\RequirePackage{tikz}
\RequirePackage{fix-cm}
\documentclass[review]{elsarticle}

\usepackage{blindtext}
\usepackage[hidelinks]{hyperref}
\usepackage[absolute,overlay]{textpos}
\usepackage[margin=1.44in]{geometry}
\usepackage{hypernat}

\usepackage{amssymb}
\usepackage{amsmath}
\usepackage{bm}
\usepackage{shellesc}
\usepackage{subcaption}
\usepackage{tikz}
\usetikzlibrary{arrows,shapes,decorations.pathreplacing,calligraphy}
\usepackage{booktabs}
\usepackage{adjustbox}
\usepackage{siunitx}
\usepackage{upgreek} 
\usepackage[utf8]{inputenc}
\usepackage[T1]{fontenc}
\usepackage{lmodern}
\usepackage{etoolbox}
\usepackage{algorithm}
\usepackage{algpseudocode}
\usepackage{siunitx}
\usepackage{physics}
\usepackage{mathdots}
\usepackage{yhmath}
\usepackage{booktabs}
\usepackage{gensymb}
\usepackage{wasysym}
\usepackage{multirow}
\usepackage{pgfplots}
\usepackage{pgfplotstable}
\pgfplotsset{compat=1.17}
\usepgfplotslibrary{polar} 
\usepackage{color}
\usepackage{multirow}
\usepackage{array}
\usepgfplotslibrary{external}
\usetikzlibrary{external}
\usetikzlibrary{fadings}
\usetikzlibrary{patterns}
\usetikzlibrary{shadows.blur}
\usetikzlibrary{arrows}
\usetikzlibrary{arrows.meta}
\usetikzlibrary{shapes}

\usepackage{algorithmicx}

\tikzset{external/system call={lualatex -shell-escape -halt-on-error 
		-interaction=batchmode -jobname "\image" "\texsource" &&
		pdftops -eps "\image.pdf" "\image.eps"}}
\usepackage[]{caption}[2008/08/24]
\definecolor{cbgreen}{RGB}{34,139,34}

\journal{Computer methods in applied mechanics and engineering}

\begin{document}

\begin{frontmatter}
	
	
	
	\title{\fontsize{13}{16}\selectfont A phase-field framework for anisotropic viscoelastic-viscoplastic fracture in short fiber-reinforced polymers in hygrothermal environments}
	%
	\author[1,2]{Behrouz Arash\corref{cor1}}
\ead{behrouza@oslomet.no}
\author[1]{Shadab Zakavati}
\author[3]{Timon Rabczuk}
\ead{timon.rabczuk@uni-weimar.de}

\address[1]{Department of Mechanical, Electrical and Chemical Engineering, Oslo Metropolitan University, Pilestredet 35, 0166 Oslo, Norway}
\address[2]{Green Energy Lab, Department of Mechanical, Electrical and Chemical Engineering, OsloMet - Oslo Metropolitan University, Oslo, Norway}
\address[3]{Institute of Structural Mechanics, Bauhaus-Universit{\"a}t Weimar, Marienstra{\ss}e 15, 99423 Weimar, Germany}

\cortext[cor1]{Corresponding author}



\begin{abstract}
	This work presents a comprehensive phase-field framework for modeling anisotropic viscoelastic-viscoplastic fracture in short fiber-reinforced polymer (SFRP) composites under hygrothermal environments at finite deformation. The constitutive model employs a multiplicative decomposition of the deformation gradient into viscoelastic and viscoplastic components. An anisotropic phase-field formulation is developed using structural tensors to capture orientation-dependent fracture energy induced by multiple fiber families. Hygrothermal effects are incorporated through moisture-dependent swelling, thermal expansion, and temperature- and moisture-sensitive material parameters within the coupled framework. Numerical investigations demonstrate the framework's capability to capture complex fracture phenomena in SFRPs. Results reveal that fiber orientation fundamentally governs the spatial distribution of crack driving force, with maximum energy accumulation along fiber directions persisting throughout viscous relaxation. The anisotropy parameter controlling directional fracture resistance significantly influences crack path deflection. Hygrothermal degradation substantially reduces both peak load and fracture energy, with moisture absorption and elevated temperature each contributing to decreased mechanical performance. The framework captures the influence of fiber mechanical properties on global load-bearing capacity and crack propagation resistance. This unified computational framework advances the predictive modeling of damage evolution in SFRPs subjected to realistic environmental and mechanical loading conditions.
\end{abstract}

\begin{keyword}
	Short fiber/Epoxy Composites \sep Hygrothermal Conditions \sep Phase-field Modeling \sep  Viscoelastic-Viscoplastic Model \sep Finite Deformation
	
	
\end{keyword}

\end{frontmatter}


\section{Introduction}
\label{sec:sec1}

Short fiber-reinforced polymer (SFRP) composites have emerged as critical materials in modern engineering applications, finding extensive use in automotive, aerospace, renewable energy, and structural engineering sectors~\cite{holbery2006natural,soutis2005carbon,arash2019viscoelastic2}. Glass fiber (GF)-reinforced epoxy composites, in particular, represent a substantial segment of the market due to their favorable cost-performance ratio and well-established processing technologies~\cite{thomason2002influence,arash2019viscoelastic2,nsengiyumva2024toward}. However, their structural integrity under service loading conditions remains critically dependent on understanding and predicting complex failure mechanisms, particularly fracture initiation and propagation under realistic environmental and mechanical loading scenarios.

The mechanical response of SFRPs exhibits pronounced anisotropy stemming from fiber orientation distributions. This microstructural anisotropy fundamentally governs not only the elastic properties but also the inelastic deformation mechanisms and fracture behavior~\cite{gusev2017finite,wang2019numerical,arash2019viscoelastic2}. Furthermore, polymer matrices demonstrate rate-dependent behavior characterized by coupled viscoelastic and viscoplastic mechanisms~\cite{poulain2014finite,arash2021finite,arash2025phase}, while hygrothermal conditions induce substantial property degradation through moisture absorption and thermal cycling~\cite{zhou2007experimental,bahtiri2023machine,arash2025phase}. The interplay between these multi-physical phenomena--mechanical anisotropy, rate-dependent inelasticity, environmental degradation, and damage evolution--presents formidable challenges for predictive modeling frameworks.

Traditional computational approaches for fracture mechanics in SFRPs have predominantly relied on cohesive zone models~\cite{ortiz1999finite,li2005use}, element deletion techniques~\cite{lapczyk2007progressive}, and extended finite element (FE) methods ~\cite{pike2015xfem,kastner2016xfem}. While these methods have demonstrated success in specific scenarios, they encounter limitations when addressing complex crack patterns, crack nucleation in initially undamaged domains, and mesh-independent crack path prediction. 
Phase-field modeling has emerged as a variational framework for fracture mechanics that fundamentally addresses limitations of classical approaches~\cite{francfort1998revisiting,bourdin2000numerical,miehe2010phase}. The phase-field method regularizes sharp crack interfaces through continuous damage field representation, thereby eliminating the need for explicit crack tracking, enabling nucleation in intact material, and providing mesh-independent solutions~\cite{ambati2015review,wu2020phase}. The approach has been successfully extended to encompass complex fracture phenomena including brittle~\cite{miehe2010thermodynamically,borden2012phase}, ductile~\cite{ambati2015phase,miehe2016phase}, and cohesive fracture~\cite{verhoosel2013phase,conti2016phase}, while recent developments have incorporated anisotropic fracture energy~\cite{li2015phase,teichtmeister2017phase,clayton2015phase}, dynamic effects~\cite{hofacker2013phase}, and multi-field coupling~\cite{wilson2013phase,miehe2015phase,arash2025phase}. 

The modeling of anisotropic fracture in fiber-reinforced composites introduces additional complexity beyond that encountered in isotropic materials. The fracture energy itself becomes directionally dependent, reflecting the preferential crack propagation paths influenced by fiber orientation~\cite{bleyer2018phase,gultekin2016phase,dean2020multi}. Several anisotropic phase-field formulations have been proposed employing structural tensors to capture orientation-dependent fracture resistance~\cite{li2015phase,teichtmeister2017phase}, while alternative approaches utilize preferential fracture directions aligned with material symmetries~\cite{clayton2015phase,bryant2018mixed}. Dean and coworkers~\cite{dean2020multi,dean2020phase,dean2025hybrid} developed multi-phase-field frameworks to simultaneously capture intralaminar matrix cracking and interlaminar delamination in laminated composites, incorporating cohesive zone behavior at ply interfaces through interfacial gradient terms. Recent advances have also addressed the multiscale nature of composite fracture, with homogenization-based phase-field models bridging microscale fiber-matrix debonding to macroscale crack propagation~\cite{nguyen2015phase,dean2020multi}. The influence of manufacturing-induced fiber orientation distributions on fracture patterns has been investigated through coupled process-structure-property phase-field frameworks~\cite{tanne2018crack,dean2020multi}, while hygrothermal degradation effects in composite phase-field models have been addressed through environment-dependent fracture energy and stiffness degradation~\cite{au2023hygroscopic,loew2020fatigue}. However, most existing anisotropic phase-field models for composites consider either elastic or small-strain elasto-plastic behavior~\cite{duda2015phase,ulloa2019phase}, with limited attention to the coupled viscoelastic-viscoplastic response characteristic of polymer matrices at finite deformation~\cite{dammass2023phase,kumar2022nonlinear}.

For polymer-based materials, several studies have developed phase-field frameworks incorporating viscoelastic behavior~\cite{yin2022viscoelastic,shen2019fracture,loew2019rate,brighenti2021phase}, recognizing the critical role of rate-dependent mechanisms in fracture processes. The time-dependent nature of polymer fracture has been captured through various approaches, including integral formulations coupled with phase-field damage~\cite{shen2019fracture}, and representative crack element frameworks for efficient time integration~\cite{yin2022viscoelastic}. Specifically for polymer nanocomposites, Arash and coworkers~\cite{arash2021finite,arash2023effect,arash2021finite2} developed phase-field formulations to investigate nonlinear viscoelastic fracture at finite deformation, accounting for hygrothermal conditions and nanoparticle reinforcement effects through moisture-dependent material parameters and thermal expansion coupling. Their work demonstrated the significance of viscoelastic relaxation in determining fracture toughness under different loading rates and environmental conditions. Recent experimental evidence indicates that epoxy resins and their composites exhibit irreversible viscoplastic strains, particularly under sustained loading or thermal cycling~\cite{poulain2014finite,BAHTIRI2023116293}. Arash et al.~\cite{arash2025phase} developed a finite deformation phase-field formulation to investigate the viscoelastic–viscoplastic fracture behavior of epoxy nanocomposites under cyclic and monolithic loading. 

However, the combination of viscoelastic and viscoplastic mechanisms within phase-field fracture frameworks remains unexplored for anisotropic SFRPs. In addition, hygrothermal effects constitute another critical consideration for polymer composite durability~\cite{liew2025experimental,arash2025phase}. Moisture absorption induces swelling, plasticization, and degradation of matrix-dominated properties, while thermal fluctuations affect glass transition behavior and time-dependent response~~\cite{botelho2006hygrothermal,arash2025phase}. Phase-field models incorporating hygro-mechanical coupling have emerged~\cite{au2023hygroscopic,arash2025phase}, though their extension to anisotropic fiber-reinforced systems with comprehensive viscoelastic-viscoplastic constitutive descriptions remains limited. Despite substantial progress in individual aspects (i.e., anisotropic phase-field modeling, viscoelastic-viscoplastic constitutive frameworks, hygrothermal coupling, and finite deformation mechanics), a comprehensive framework unifying these elements for SFRPs remains absent from the literature. 

Motivated by these considerations, the present work develops an integrated phase-field framework for modeling anisotropic viscoelastic-viscoplastic fracture in SFRPs under hygrothermal environments at finite deformation. The principal contributions and novel aspects of this study encompass several key developments. First, a comprehensive constitutive model is formulated that decomposes the deformation into viscoelastic and viscoplastic components, thereby extending the framework of Arash et al.~\cite{arash2021finite,arash2023effect} to incorporate viscoplastic flow. Second, an anisotropic phase-field fracture formulation is developed employing structural tensors to capture orientation-dependent fracture energy induced by multiple fiber families, enabling the prediction of preferential crack paths in layered and randomly distributed fiber architectures. Third, hygrothermal effects are integrated through moisture-dependent swelling, thermal expansion, and temperature-sensitive material parameters within the coupled viscoelastic-viscoplastic-damage framework. Finally, systematic numerical investigations are conducted demonstrating the model's capability to capture the influence of fiber orientation, distribution and performance on the fracture behavior, the role of anisotropic fracture energy in crack deflection, and hygrothermal degradation effects on fracture resistance.

The remainder of this paper is organized as follows. Section~\ref{sec:const} presents the viscoelastic-viscoplastic constitutive model for SFRPs, detailing the kinematics, stress decomposition, fiber family representation, and hygrothermal coupling. Sec.~\ref{sec:phase} introduces the anisotropic phase-field formulation. Sec.~\ref{sec:results} demonstrates the model through representative numerical examples examining the effects of fiber orientation, anisotropy parameters, and hygrothermal conditions on fracture evolution. Finally, Section~\ref{sec:summary} summarizes the key findings and discusses future research directions.

\section{Constitutive model for short fiber-reinforced epoxy composites}
\label{sec:const}

\begin{figure}[!ht]
\centering
\begin{subfigure}[b]{0.49\textwidth}
	\centering
	\includegraphics{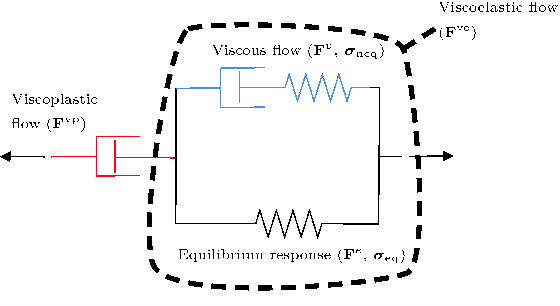}
	\caption{}
	\label{fig:rheological_model_a}
\end{subfigure}%
\hfill
\begin{subfigure}[b]{0.49\textwidth}
	\centering
	\includegraphics{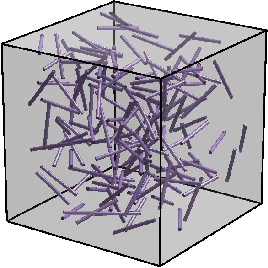}
	\caption{}	
	\label{fig:rheological_model_b}
\end{subfigure}
\caption{(a) One-dimensional schematic of the viscoelastic-viscoplastic constitutive model, and (b) representative volume element of SFRPs.}
\label{fig:rheological_model}
\end{figure}

In this section, a viscoelastic-viscoplastic constitutive model is presented for SFRPs at finite deformation. The model captures the nonlinear rate-dependent behavior of the matrix phase combined with the anisotropic reinforcement provided by multiple fiber families. The constitutive framework also includes hygrothermal effects. A schematic representation of the rheological model is shown in Fig.~\ref{fig:rheological_model}. In the current formulation, temperature $\theta$ and moisture content $w_w$ are treated as prescribed uniform parameters rather than independent field variables. Governing equations for heat transfer or moisture diffusion are not solved; instead, spatially uniform and temporally constant values are assumed, representing steady-state environmental conditions. This approach captures the essential hygrothermal degradation mechanisms while maintaining computational tractability.

\subsection{Kinematics}

The total deformation gradient is multiplicatively decomposed into volumetric and deviatoric parts as
\begin{equation}
\mathbf{F} = J^{1/3}\bar{\mathbf{F}},
\label{eq:F_decomp}
\end{equation}
where $J = \det[\mathbf{F}]$ and $\bar{\mathbf{F}}$ represent the volumetric and deviatoric deformation gradients, respectively. The volumetric deformation is further decomposed to account for mechanical compressibility, thermal dilatation, and moisture-induced swelling as
\begin{equation}
J = J_m J_\theta J_w,
\label{eq:J_decomp}
\end{equation}
where
\begin{equation}
J_\theta = 1 + \alpha_\theta (\theta - \theta_0),
\label{eq:J_theta}
\end{equation}
and
\begin{equation}
J_w = 1 + \alpha_w w_w.
\label{eq:J_w}
\end{equation}

Here, $\alpha_\theta$ and $\alpha_w$ denote the thermal expansion and moisture swelling coefficients~\cite{arash2023effect}, respectively, $\theta$ is the absolute temperature, $\theta_0 = 296$ K is the reference temperature, and $w_w$ represents the moisture content.

The material response is decomposed into viscoelastic and viscoplastic components. The viscoelastic behavior is further split into equilibrium (hyperelastic) and non-equilibrium (viscous) contributions. Accordingly, the deviatoric deformation gradient is multiplicatively decomposed as~\cite{govindjee1997presentation}
\begin{equation}
\bar{\mathbf{F}} = \bar{\mathbf{F}}^{\text{ve}}\bar{\mathbf{F}}^{\text{vp}},
\label{eq:Fbar_decomp}
\end{equation}
where $\bar{\mathbf{F}}^{\text{ve}}$ and $\bar{\mathbf{F}}^{\text{vp}}$ represent the viscoelastic and viscoplastic deformation gradients, respectively. The viscoelastic deformation gradient is further decomposed into elastic and viscous parts as
\begin{equation}
\bar{\mathbf{F}}^{\text{ve}} = \bar{\mathbf{F}}^e\bar{\mathbf{F}}^v.
\label{eq:Fve_decomp}
\end{equation}

The corresponding left Cauchy-Green deformation tensors are defined as
\begin{align}
\bar{\mathbf{B}} &= \bar{\mathbf{F}}\bar{\mathbf{F}}^T, \label{eq:Bbar}\\
\bar{\mathbf{B}}^e &= \bar{\mathbf{F}}^e(\bar{\mathbf{F}}^e)^T, \label{eq:Be}\\
\bar{\mathbf{B}}^{\text{ve}} &= \bar{\mathbf{F}}^{\text{ve}}(\bar{\mathbf{F}}^{\text{ve}})^T. \label{eq:Bve}
\end{align}

The total velocity gradient of the viscoelastic network, $\bar{\mathbf{L}}^{\text{ve}} = \dot{\bar{\mathbf{F}}}^{\text{ve}}(\bar{\mathbf{F}}^{\text{ve}})^{-1}$, is additively decomposed into elastic and viscous components:
\begin{equation}
\bar{\mathbf{L}}^{\text{ve}} = \bar{\mathbf{L}}^e + \bar{\mathbf{F}}^e\bar{\mathbf{L}}^v(\bar{\mathbf{F}}^e)^{-1} = \bar{\mathbf{L}}^e + \tilde{\mathbf{L}}^v,
\label{eq:Lve_decomp}
\end{equation}
where
\begin{equation}
\tilde{\mathbf{L}}^v = \dot{\bar{\mathbf{F}}}^v(\bar{\mathbf{F}}^v)^{-1} = \tilde{\mathbf{D}}^v + \tilde{\mathbf{W}}^v.
\label{eq:Lv_decomp}
\end{equation}

Here, a tilde ($\sim$) denotes quantities in the intermediate configuration, $\tilde{\mathbf{D}}^v$ represents the rate of viscous deformation, and $\tilde{\mathbf{W}}^v$ is the skew-symmetric spin tensor. Following standard practice \cite{qi2005stress}, the intermediate configuration is made unique by prescribing $\tilde{\mathbf{W}}^v = \mathbf{0}$.

The rate of viscoelastic flow is described \cite{boyce1988large} as
\begin{equation}
\tilde{\mathbf{D}}^v = \frac{\dot{\varepsilon}^v}{\tau_{\text{neq}}} \text{dev}\langle\boldsymbol{\sigma}'_{\text{neq}}\rangle,
\label{eq:Dv}
\end{equation}
where $\tau_{\text{neq}} = \|\text{dev}[\boldsymbol{\sigma}_{\text{neq}}]\|_F$ represents the Frobenius norm of the driving stress, $\dot{\varepsilon}^v$ is the viscous strain rate, and $\boldsymbol{\sigma}'_{\text{neq}} = \mathbf{R}_e^T\boldsymbol{\sigma}_{\text{neq}}\mathbf{R}_e$ represents the stress acting on the viscous component in its relaxed configuration. The viscous strain rate is defined by the Argon model~\cite{arruda1993three} as
\begin{equation}
\dot{\varepsilon}^v = \dot{\varepsilon}_0 \exp\left[\frac{\Delta H}{k_b T}\left(\left(\frac{\tau_{\text{neq}}}{\tau_0}\right)^m - 1\right)\right],
\label{eq:eps_dot_v}
\end{equation}
where $k_b$ is the Boltzmann constant, $\dot{\varepsilon}_0$ is a pre-exponential factor, $\Delta H$ is the activation energy, $\tau_0$ is the athermal yield stress, and $m$ is an exponential material parameter that captures the moisture- or temperature-dependency of the stiffness~\cite{bahtiri2023machine}.

From Eqs.~\eqref{eq:Lv_decomp} and \eqref{eq:Dv}, the evolution of the viscous deformation gradient is given by
\begin{equation}
\dot{\bar{\mathbf{F}}}^v = (\bar{\mathbf{F}}^e)^{-1}\frac{\dot{\varepsilon}^v}{\tau_{\text{neq}}}\text{dev}\langle\boldsymbol{\sigma}'_{\text{neq}}\rangle\bar{\mathbf{F}}^{\text{ve}}.
\label{eq:Fv_evol}
\end{equation}

Similarly, the total velocity gradient of the overall network, $\bar{\mathbf{L}} = \dot{\bar{\mathbf{F}}}(\bar{\mathbf{F}})^{-1}$, can be expanded to
\begin{equation}
\bar{\mathbf{L}} = \bar{\mathbf{L}}^{\text{ve}} + \bar{\mathbf{F}}^{\text{ve}}\bar{\mathbf{L}}^{\text{vp}}(\bar{\mathbf{F}}^{\text{ve}})^{-1} = \bar{\mathbf{L}}^{\text{ve}} + \tilde{\mathbf{L}}^{\text{vp}}.
\label{eq:L_decomp}
\end{equation}

The viscoplastic velocity gradient is additively decomposed into symmetric and skew-symmetric parts as
\begin{equation}
\tilde{\mathbf{L}}^{\text{vp}} = \dot{\bar{\mathbf{F}}}^{\text{vp}}(\bar{\mathbf{F}}^{\text{vp}})^{-1} = \tilde{\mathbf{D}}^{\text{vp}} + \tilde{\mathbf{W}}^{\text{vp}},
\label{eq:Lvp_decomp}
\end{equation}
with $\tilde{\mathbf{W}}^{\text{vp}} = \mathbf{0}$. The rate of viscoplastic deformation is defined by
\begin{equation}
\tilde{\mathbf{D}}^{\text{vp}} = \frac{\dot{\varepsilon}^{\text{vp}}}{\tau_{\text{tot}}} \text{dev}[\boldsymbol{\sigma}],
\label{eq:Dvp}
\end{equation}
where $\text{dev}[\boldsymbol{\sigma}]$ is the deviatoric part of the total Cauchy stress and $\tau_{\text{tot}} = \|\text{dev}[\boldsymbol{\sigma}]\|_F$.

To characterize the viscoplastic flow, a simple phenomenological representation is adopted as
\begin{equation}
\dot{\varepsilon}^{\text{vp}} = 
\begin{cases}
	0 & \text{if } \tau_{\text{tot}} < \sigma_0, \\
	a(\varepsilon - \varepsilon_0)^b \dot{\varepsilon} & \text{if } \tau_{\text{tot}} \geq \sigma_0,
\end{cases}
\label{eq:eps_dot_vp}
\end{equation}
where $a$, $b$, and $\sigma_0$ are material parameters, $\varepsilon_0$ is the strain at which viscoplastic flow is activated, represented by the Frobenius norm of the Green strain tensor $\varepsilon = \|\mathbf{E}\|_F$ with
\begin{equation}
\mathbf{E} = \frac{1}{2}(\mathbf{F}^T\mathbf{F} - \mathbf{I}),
\label{eq:Green_strain}
\end{equation}
and $\dot{\varepsilon}$ is the strain rate, introducing a simple strain-rate dependency of the viscoplastic flow.

Analogous to Eq.~\eqref{eq:Fv_evol}, the evolution of the viscoplastic deformation gradient is
\begin{equation}
\dot{\bar{\mathbf{F}}}^{\text{vp}} = (\bar{\mathbf{F}}^{\text{ve}})^{-1}\frac{\dot{\varepsilon}^{\text{vp}}}{\tau_{\text{tot}}}\text{dev}[\boldsymbol{\sigma}_{\text{tot}}]\bar{\mathbf{F}}.
\label{eq:Fvp_evol}
\end{equation}

\subsection{Implicit time integration}
\label{sec:time_integration}

The evolution equations~\eqref{eq:Fv_evol} and~\eqref{eq:Fvp_evol} for $\bar{\mathbf{F}}^v$
and $\bar{\mathbf{F}}^{vp}$ are first-order nonlinear ordinary differential equations whose right-hand sides depend on the Cauchy stress at the current time, which is itself a function of the unknown internal variables at $t_{n+1}$.
Two numerical challenges therefore arise. First, naive additive updates of the form
$\bar{\mathbf{F}}^{v}_{n+1} = \bar{\mathbf{F}}^{v}_{n} + \Delta t\,\dot{\bar{\mathbf{F}}}^{v}$
do not preserve the isochoric constraints
$\det(\bar{\mathbf{F}}^{v}) = \det(\bar{\mathbf{F}}^{vp}) = 1$. Second, the implicit dependence of the viscous and viscoplastic flow directions on the unknown stress at $t_{n+1}$ requires an iterative solution within each time step.

Both issues are resolved by adopting an exponential map integrator for the deviatoric internal deformation gradients. For a generic deviatoric velocity gradient $\bar{\mathbf{L}}$ (either $\bar{\mathbf{L}}^{v}$ or $\bar{\mathbf{L}}^{vp}$), the update reads
\begin{equation}
	\bar{\mathbf{F}}_{n+1}
	= \exp\!\bigl(\Delta t\,\bar{\mathbf{L}}_{n+1}\bigr)\,\bar{\mathbf{F}}_{n}.
	\label{eq:exp_map}
\end{equation}
Because the flow rules prescribe traceless velocity gradients,
$\operatorname{tr}(\bar{\mathbf{L}}^{v}) = \operatorname{tr}(\bar{\mathbf{L}}^{vp}) = 0$, the matrix exponential satisfies $\det\!\bigl(\exp(\Delta t\,\bar{\mathbf{L}})\bigr)
= \exp\!\bigl(\Delta t\,\operatorname{tr}\bar{\mathbf{L}}\bigr) = 1$. Accordingly, The isochoric constraint is preserved exactly at every iteration.

Since $\bar{\mathbf{L}}_{n+1}$ depends on the stress at $t_{n+1}$, the update in Eq.~\eqref{eq:exp_map} is implicit. We solve it by means of a fixed-point iteration in the spirit of return-mapping algorithms~\cite{simo1992algorithms}. This procedure is detailed in Algorithm~\ref{alg:time_integration}. The rotation tensor $\mathbf{R}_{e}$ appearing in the algorithm is obtained from the polar decomposition of the elastic part of the deviatoric deformation gradient,
\begin{equation}
	\bar{\mathbf{F}}^{e} = \mathbf{R}_{e}\,\mathbf{U}_{e},
	\label{eq:polar_dec}
\end{equation}
where $\mathbf{R}_{e}$ is proper orthogonal and $\mathbf{U}_{e}$ is symmetric positive definite.

\begin{algorithm}[H]
	\caption{Implicit time integration over $[t_n,t_{n+1}]$ with fixed‑point iteration.}
	\label{alg:time_integration}
	\begin{algorithmic}[1]
		\State \textbf{Given:} $\mathbf{F}_{n+1}$, $\bar{\mathbf{F}}^v_n$, $\bar{\mathbf{F}}^{vp}_n$, $\mathcal{H}_n$, time step $\Delta t$.
		\State \textbf{Initialize:} $\bar{\mathbf{F}}^v = \bar{\mathbf{F}}^v_n$, $\bar{\mathbf{F}}^{vp} = \bar{\mathbf{F}}^{vp}_n$, tolerance $\varepsilon_{tol}$, $\texttt{converged}= \text{false}$.
		\While{$\texttt{converged}= \text{false}$}
		\State \textbf{Kinematics:}
		$J = \det(\mathbf{F}_{n+1})$, $\bar{\mathbf{F}} = J^{-1/3}\mathbf{F}_{n+1}$,
		$\bar{\mathbf{F}}^{ve} = \bar{\mathbf{F}}\,(\bar{\mathbf{F}}^{vp})^{-1}$,
		$\bar{\mathbf{F}}^{e} = \bar{\mathbf{F}}^{ve}\,(\bar{\mathbf{F}}^v)^{-1}$.
		\State \textbf{Stress:} Compute Cauchy stress $\bm{\sigma}$ using the constitutive model (Section~\ref{subsec:Helmholtz}).
		\State \textbf{Polar decomposition:} $\bar{\mathbf{F}}^{e} = \mathbf{R}_e\mathbf{U}_e$ with $\mathbf{R}_e$ orthogonal, $\mathbf{U}_e$ symmetric.
		\State \textbf{Rotated non‑equilibrium stress:} $\bm{\sigma}_{\mathrm{neq}}' = \mathbf{R}_e^T\,\bm{\sigma}_{\mathrm{neq}}\,\mathbf{R}_e$.
		\State \textbf{Viscous flow:}
		$\tau_{\mathrm{neq}} = \|\operatorname{dev}[\bm{\sigma}_{\mathrm{neq}}]\|_F$,
		$\dot{\epsilon}^v$ from Eq.~\eqref{eq:eps_dot_v},
		$\bar{\mathbf{D}}^v = \dfrac{\dot{\epsilon}^v}{\tau_{\mathrm{neq}}}\operatorname{dev}[\bm{\sigma}_{\mathrm{neq}}']$,
		$\bar{\mathbf{L}}^v = \bar{\mathbf{D}}^v$ (irrotational flow),
		$\bar{\mathbf{F}}^v_{\text{new}} = \exp\!\big(\Delta t\,\bar{\mathbf{L}}^v\big)\,\bar{\mathbf{F}}^v$.
		\State \textbf{Viscoplastic flow:}
		$\tau_{\mathrm{tot}} = \|\operatorname{dev}[\bm{\sigma}]\|_F$,
		$\mathbf{E} = \tfrac12(\mathbf{F}_{n+1}^T\mathbf{F}_{n+1}-\mathbf{I})$, $\epsilon = \|\mathbf{E}\|_F$,
		\If{$\tau_{\mathrm{tot}} \ge \sigma_0$}
		\State $\dot{\epsilon}^{vp}$ from Eq.~\eqref{eq:eps_dot_vp},
		$\bar{\mathbf{D}}^{vp} = \dfrac{\dot{\epsilon}^{vp}}{\tau_{\mathrm{tot}}}\operatorname{dev}[\bm{\sigma}]$,
		$\bar{\mathbf{L}}^{vp} = \bar{\mathbf{D}}^{vp}$,
		$\bar{\mathbf{F}}^{vp}_{\text{new}} = \exp\!\big(\Delta t\,\bar{\mathbf{L}}^{vp}\big)\,\bar{\mathbf{F}}^{vp}$.
		\Else		\State $\bar{\mathbf{F}}^{vp}_{\text{new}} = \bar{\mathbf{F}}^{vp}$.
		\EndIf
		\State \textbf{Convergence check:}
		$\Delta^v = \|\bar{\mathbf{F}}^v_{\text{new}} - \bar{\mathbf{F}}^v\|_F$,
		$\Delta^{vp}= \|\bar{\mathbf{F}}^{vp}_{\text{new}} - \bar{\mathbf{F}}^{vp}\|_F$.
		\If{$\Delta^v + \Delta^{vp} < \varepsilon_{tol}$}
		\State $\bar{\mathbf{F}}^v_{n+1} = \bar{\mathbf{F}}^v_{\text{new}}$, $\bar{\mathbf{F}}^{vp}_{n+1} = \bar{\mathbf{F}}^{vp}_{\text{new}}$, $\texttt{converged}= \text{true}$.
		\Else		\State $\bar{\mathbf{F}}^v = \bar{\mathbf{F}}^v_{\text{new}}$, $\bar{\mathbf{F}}^{vp} = \bar{\mathbf{F}}^{vp}_{\text{new}}$.
		\EndIf
		\EndWhile
	\end{algorithmic}
\end{algorithm}

It is required to rotate the non-equilibrium Cauchy stress $\bm{\sigma}_{\mathrm{neq}}$ to the intermediate (stress-free) configuration, yielding
\begin{equation}
	\bm{\sigma}'_{\mathrm{neq}}
	= \mathbf{R}_{e}^{T}\,\bm{\sigma}_{\mathrm{neq}}\,\mathbf{R}_{e},
	\label{eq:rotated_neq}
\end{equation}
consistent with the objective flow rule in Eq.~\eqref{eq:Dv} and with the requirement that the intermediate configuration be made unique by setting the viscous spin to zero~\cite{reese1998theory}. The algorithm guarantees exact satisfaction of the isochoric constraints
$\det(\bar{\mathbf{F}}^{v}_{n+1}) = \det(\bar{\mathbf{F}}^{vp}_{n+1}) = 1$ at convergence, and produces a stress state that is fully consistent with the viscous and viscoplastic evolution equations at $t_{n+1}$.

\subsection{Helmholtz free energy and stress response}
\label{subsec:Helmholtz}

Following the additive decomposition proposed in previous work \cite{ambati2016phase}, the Helmholtz free energy is decomposed into equilibrium, non-equilibrium, and volumetric contributions as
\begin{equation}
\psi(\bar{\mathbf{B}}^{\text{ve}}, \bar{\mathbf{B}}^e, J, \phi) = g(\phi)\left[\psi_{\text{eq}}(\bar{\mathbf{B}}^{\text{ve}}) + \psi_{\text{neq}}(\bar{\mathbf{B}}^e)\right] + \psi_{\text{vol}}(J),
\label{eq:free_energy}
\end{equation}
where $g(\phi)$ is the energetic degradation function capturing the evolution of strain energy with respect to the phase-field variable $\phi \in [0,1]$. The degradation function satisfies
\begin{equation}
g(0) = 1, \quad g(1) = 0, \quad g'(\phi) \leq 0, \quad g'(1) = 0.
\label{eq:degradation_conditions}
\end{equation}

A monotonically decreasing function satisfying the conditions in Eq.~\eqref{eq:degradation_conditions} is adopted as
\begin{equation}
g(\phi) = (1 - \phi)^2 + k,
\label{eq:degradation_function}
\end{equation}
where $k$ is a small positive parameter introduced to ensure numerical stability and prevent complete loss of stiffness at fully damaged material points~\cite{miehe2010thermodynamically}.

For SFRPs, the equilibrium and non-equilibrium parts of the free energy account for both matrix and fiber contributions. The strain energy density for each fiber family is based on a hyperelastic model that captures the anisotropic mechanical response. The equilibrium and non-equilibrium free energies are defined~\cite{arash2019viscoelastic2} as
\begin{equation}
\rho_0\psi_{\text{eq}} = \sum_{i=1}^{N_f} v_f^{(i)}\left[\psi_{\text{eq}}^{\text{matrix}}(\bar{\mathbf{B}}^{\text{ve}}) + \psi_{\text{eq}}^{\text{fiber}(i)}(\bar{\mathbf{C}}^{\text{ve}}, \mathbf{a}_0^{(i)})\right],
\label{eq:psi_eq}
\end{equation}
\begin{equation}
\rho_0\psi_{\text{neq}} = \sum_{i=1}^{N_f} v_f^{(i)}\left[\psi_{\text{neq}}^{\text{matrix}}(\bar{\mathbf{B}}^e) + \psi_{\text{neq}}^{\text{fiber}(i)}(\bar{\mathbf{C}}^e, \mathbf{a}_0^{(i)})\right],
\label{eq:psi_neq}
\end{equation}
where $N_f$ is the number of fiber families, $v_f^{(i)}$ is the volume fraction of the $i$-th fiber family, $\mathbf{a}_0^{(i)}$ is the unit vector defining the fiber direction in the reference configuration, $\bar{\mathbf{C}}^{\text{ve}} = (\bar{\mathbf{F}}^{\text{ve}})^T\bar{\mathbf{F}}^{\text{ve}}$ and $\bar{\mathbf{C}}^e = (\bar{\mathbf{F}}^e)^T\bar{\mathbf{F}}^e$ are the right Cauchy-Green deformation tensors, and $\rho_0$ is the initial density. The free energy density is expressed using both left and right Cauchy-Green tensors to reflect the different physical origins of the contributions. The matrix material, being isotropic, is naturally described using the left Cauchy-Green tensor $\bar{\mathbf{B}}$ in the current configuration. The fiber reinforcement, with its directional dependence defined in the reference configuration, requires the right Cauchy-Green tensor $\bar{\mathbf{C}}$ because the squared fiber stretch is correctly given by $\bar{I}_4^{(i)} = \mathbf{a}_0^{(i)}\cdot\bar{\mathbf{C}}\mathbf{a}_0^{(i)}$.

The matrix contribution is modeled using a neo-Hookean hyperelastic model as
\begin{equation}
\psi_{\text{eq/neq}}^{\text{matrix}} = \frac{1}{2}\mu_{\text{eq/neq}}(\theta, w_w)(\text{tr}[\bar{\mathbf{B}}^{\text{ve/e}}] - 3),
\label{eq:psi_matrix}
\end{equation}
where $\mu_{\text{eq}}$ and $\mu_{\text{neq}}$ are temperature- and moisture-dependent shear moduli. The temperature dependence is captured using a modified Kitagawa model \cite{unger2020effect}:
\begin{align}
\mu_{\text{eq}}(\theta, w_w) &= \mu_{\text{eq}}^0[2 - \exp(\alpha(\theta - \theta_0))][1 - 9.5w_w + 0.057w_w^2], \label{eq:mu_eq}\\
\mu_{\text{neq}}(\theta, w_w) &= \mu_{\text{neq}}^0[2 - \exp(\alpha(\theta - \theta_0))][1 - 9.5w_w + 0.057w_w^2], \label{eq:mu_neq}
\end{align}
where $\mu_{\text{eq}}^0$ and $\mu_{\text{neq}}^0$ are reference shear moduli at room temperature and dry conditions, and $\alpha$ is a temperature sensitivity parameter.

The fiber contribution is based on a strain energy function that accounts for fiber stretching and shearing. The fiber energy for each family is expressed in terms of fiber-specific invariants as
\begin{align}
\psi_{\text{fiber}}^{(i)} = &\frac{1}{2}\mu\bigg[(v_m + v_f^{(i)}f(\bar{I}_4^{(i)}))(\bar{I}_4^{(i)} + 2(\bar{I}_4^{(i)})^{-1/2} - 3) \nonumber\\ &+ g_1(\bar{I}_4^{(i)})(\bar{I}_5^{(i)} - (\bar{I}_4^{(i)})^2)\bar{I}_4^{(i)}{}^{-1} + g_2(\bar{I}_4^{(i)})(\bar{I}_1 - (\bar{I}_5^{(i)} + 2(\bar{I}_4^{(i)})^{1/2})\bar{I}_4^{(i)}{}^{-1})\bigg],
\label{eq:psi_fiber}
\end{align}
where $v_m = 1 - \sum_{i=1}^{N_f} v_f^{(i)}$ is the matrix volume fraction, $\mu$ is the shear modulus (either $\mu_{\text{eq}}$ or $\mu_{\text{neq}}$), and the fiber-specific invariants are
\begin{align}
\bar{I}_4^{(i)} &= \mathbf{a}_0^{(i)} \cdot \bar{\mathbf{C}} \mathbf{a}_0^{(i)}, \label{eq:I4}\\
\bar{I}_5^{(i)} &= \mathbf{a}_0^{(i)} \cdot \bar{\mathbf{C}}^2 \mathbf{a}_0^{(i)}, \label{eq:I5}
\end{align}
representing the squared fiber stretch and a higher-order fiber invariant, respectively.

The functions $f$, $g_1$, and $g_2$ in Eq.~\eqref{eq:psi_fiber} are defined as
\begin{equation}
f(\bar{I}_4) = a_1 + a_2\exp[a_3(\bar{I}_4 - 1)],
\label{eq:f_function}
\end{equation}
\begin{equation}
g_1(\bar{I}_4) = \frac{(1 + v_f^{(i)})f(\bar{I}_4) + (1 - v_f^{(i)})}{(1 - v_f^{(i)})f(\bar{I}_4) + 1 + v_f^{(i)}},
\label{eq:g1_function}
\end{equation}
\begin{equation}
g_2(\bar{I}_4) = \frac{(1 + 0.4v_f^{(i)})f(\bar{I}_4) + 0.4(1 - v_f^{(i)})}{(1 - v_f^{(i)})f(\bar{I}_4) + 0.4 + v_f^{(i)}},
\label{eq:g2_function}
\end{equation}
where $a_1$, $a_2$, and $a_3$ are material parameters characterizing the fiber response.

The volumetric part of the free energy is defined as
\begin{equation}
\rho_0\psi_{\text{vol}} = \frac{1}{2}k_v\left(\frac{J_m^2 - 1}{2} - \ln[J_m]\right),
\label{eq:psi_vol}
\end{equation}
where the bulk modulus $k_v$ is temperature- and moisture-dependent given by
\begin{equation}
k_v = k_v^0[2 - \exp(\alpha(\theta - \theta_0))][1 - 9.5w_w + 0.057w_w^2],
\label{eq:kv}
\end{equation}
with $k_v^0$ being the reference bulk modulus.

The Cauchy stress is obtained from the free energy as
\begin{equation}
\boldsymbol{\sigma} = g(\phi)(\boldsymbol{\sigma}_{\text{dev}} + \boldsymbol{\sigma}_{\text{vol}}),
\label{eq:sigma_total}
\end{equation}
where the deviatoric stress is
\begin{equation}
\boldsymbol{\sigma}_{\text{dev}} = J^{-1}\sum_{i=1}^{N_f} v_f^{(i)}\left[\boldsymbol{\sigma}_{\text{matrix}} + \boldsymbol{\sigma}_{\text{fiber}}^{(i)}\right],
\label{eq:sigma_dev}
\end{equation}
and the volumetric stress is
\begin{equation}
\boldsymbol{\sigma}_{\text{vol}} = \frac{1}{2}k_v J^{-1}\left(J_m - \frac{1}{J_m}\right)\mathbf{I}.
\label{eq:sigma_vol}
\end{equation}

The matrix contribution to the deviatoric stress follows from the neo-Hookean model as
\begin{equation}
\boldsymbol{\sigma}_{\text{matrix}} = \mu_{\text{eq}}\text{dev}[\bar{\mathbf{B}}^{\text{ve}}] + \mu_{\text{neq}}\text{dev}[\bar{\mathbf{B}}^e],
\label{eq:sigma_matrix}
\end{equation}
where $\text{dev}[\cdot]$ denotes the deviatoric part.

The fiber contribution to the stress for each family is derived from the strain energy function and involves derivatives with respect to the fiber invariants. The fiber stress contribution is
\begin{align}
\boldsymbol{\sigma}_{\text{fiber}}^{(i)} = &\frac{2}{J}\bigg[W_1^{(i)}\text{dev}[\bar{\mathbf{B}}] + W_4^{(i)}\bar{I}_4^{(i)}(\mathbf{a} \otimes \mathbf{a} - \frac{1}{3}\mathbf{I}) \nonumber\\
&+ W_5^{(i)}\bar{I}_4^{(i)}(\mathbf{a} \otimes \bar{\mathbf{B}}\mathbf{a} + \bar{\mathbf{B}}\mathbf{a} \otimes \mathbf{a} - \frac{2}{3}\bar{I}_5^{(i)}\mathbf{I})\bigg],
\label{eq:sigma_fiber}
\end{align}
where $\mathbf{a} = \bar{\mathbf{F}}\mathbf{a}_0^{(i)}/(J^{2/3}\bar{I}_4^{(i)})^{1/2}$ is the current fiber direction, and the coefficients $W_1^{(i)}$, $W_4^{(i)}$, and $W_5^{(i)}$ are
\begin{equation}
W_1^{(i)} = \frac{1}{2}\mu g_2,
\label{eq:W1}
\end{equation}
\begin{align}
W_4^{(i)} = &\frac{1}{2}\mu\bigg[v_f^{(i)}f'(\bar{I}_4 + 2(\bar{I}_4)^{-1/2} - 3) + (v_m + v_f^{(i)}f)(1 - (\bar{I}_4)^{-3/2}) \nonumber\\
&- g_1(\bar{I}_5(\bar{I}_4)^{-2} + 1) + g_2(\bar{I}_5(\bar{I}_4)^{-2} + (\bar{I}_4)^{-3/2}) \nonumber\\
&+ \frac{(\bar{I}_5 - (\bar{I}_4)^2)}{2\bar{I}_4}g_1' + \frac{1}{2}\left(\bar{I}_1 - \frac{\bar{I}_5 + 2(\bar{I}_4)^{1/2}}{\bar{I}_4}\right)g_2'\bigg],
\label{eq:W4}
\end{align}
\begin{equation}
W_5^{(i)} = \frac{\mu}{2\bar{I}_4}(g_1 - g_2),
\label{eq:W5}
\end{equation}
where primes denote derivatives with respect to $\bar{I}_4^{(i)}$.

\subsection{Fiber orientation and multiple fiber families}

In short fiber-reinforced composites, fibers are typically distributed with varying orientations throughout the material. To characterize this orientation distribution in a computationally tractable manner, the fiber microstructure is represented using a second-order orientation tensor $\mathbf{A}$.

The second-order orientation tensor is constructed by averaging over all individual fibers in the representative volume element as
\begin{equation}
\mathbf{A} = \sum_{k=1}^{N_{\text{fibers}}} w_k \, \mathbf{a}_k \otimes \mathbf{a}_k,
\label{eq:orientation_tensor}
\end{equation}
where $N_{\text{fibers}}$ is the total number of fibers in the RVE, $\mathbf{a}_k$ is the unit vector along the direction of the $k$-th fiber, and $w_k = 1/N_{\text{fibers}}$ are equal weighting factors such that $\sum_{k=1}^{N_{\text{fibers}}} w_k = 1$. This tensor represents the homogenized internal orientation state of the material and captures the anisotropy induced by the fiber distribution.

The orientation tensor $\mathbf{A}$ is symmetric and positive semi-definite, with its eigenvalues providing information about the degree of alignment along different spatial directions. For a fully random (isotropic) fiber distribution in 3D, $\mathbf{A} = \frac{1}{3}\mathbf{I}$, whereas for perfectly aligned fibers in a single direction $\mathbf{n}$, $\mathbf{A} = \mathbf{n} \otimes \mathbf{n}$.

To incorporate the fiber orientation into the constitutive model, the orientation tensor is decomposed using eigenvalue decomposition as
\begin{equation}
\mathbf{A} = \sum_{i=1}^{n_{\text{dim}}} \lambda_i \, \mathbf{n}_i \otimes \mathbf{n}_i,
\label{eq:eigendecomposition}
\end{equation}
where $\lambda_i$ are the eigenvalues, $\mathbf{n}_i$ are the corresponding orthonormal eigenvectors, and $n_{\text{dim}}$ is the spatial dimension (2 for plane problems, 3 for general 3D problems). The eigenvalues satisfy $\sum_{i=1}^{n_{\text{dim}}} \lambda_i = 1$ and represent the relative concentration of fibers along each principal direction.

The eigenvectors $\mathbf{n}_i$ define the principal fiber directions, which represent the dominant orientational axes in the material. This decomposition effectively reduces the complex multi-fiber orientation distribution to a small number of principal fiber families, each characterized by a principal direction: $\mathbf{a}^{(i)}_0 = \mathbf{n}_i / \|\mathbf{n}_i\|$ and an orientation parameter, $\lambda_i$.

Given the total fiber volume fraction $v_f$ in the composite and the eigenvalues from the orientation tensor decomposition, the volume fraction assigned to each principal fiber family is determined as
\begin{equation}
v^{(i)}_f = v_f \cdot \frac{\lambda_i}{\sum_{j=1}^{N_f} \lambda_j},
\label{eq:fiber_volume_fraction}
\end{equation}
where $v^{(i)}_f$ represents the effective volume fraction of the $i$-th fiber family. This distribution ensures that
\begin{equation}
\sum_{i=1}^{N_f} v^{(i)}_f = v_f,
\label{eq:volume_conservation}
\end{equation}
preserving the total fiber content in the material.

The matrix volume fraction is then computed as
\begin{equation}
v_m = 1 - \sum_{i=1}^{N_f} v^{(i)}_f = 1 - v_f.
\label{eq:matrix_volume_fraction}
\end{equation}

This fiber family decomposition has several important implications for the constitutive model. First, the material response exhibits directional dependence, with enhanced stiffness and strength along principal fiber directions having larger eigenvalues $\lambda_i$. Second, rather than explicitly tracking thousands of individual fibers with different orientations, the model captures the essential anisotropy through a small number of principal families (2 or 3), significantly reducing computational cost while preserving the key mechanical characteristics. Third, the framework can represent any fiber orientation distribution, from fully random states to highly aligned configurations (where one eigenvalue approaches 1 and others approach 0), or intermediate states with partial alignment. Finally, the total strain energy density is computed as a weighted sum over fiber families according to Eqs.~\eqref{eq:psi_eq} and \eqref{eq:psi_neq}, where each family contributes proportionally to its volume fraction $v^{(i)}_f$. The orientation tensor approach thus provides a rigorous homogenization framework that bridges the microscopic fiber distribution and the macroscopic anisotropic constitutive behavior, enabling accurate prediction of the directionally dependent mechanical response and fracture characteristics of short fiber-reinforced polymer composites.

\section{Phase-field model at finite deformation}
\label{sec:phase}
To evaluate the predictive capability of the proposed constitutive model, we incorporate it within a phase-field fracture framework for short fiber-reinforced polymer composites. This section presents the variational phase-field formulation for quasi-brittle fracture at finite deformation, along with the continuum mechanics incremental scheme and FE equations.

\subsection{Problem field description}

The strong form of the boundary value problem in the spatial form for the coupled displacement $\mathbf{u}$ and phase-field variable $\phi$ can be written as
\begin{align}
	\nabla_{\mathbf{x}} \cdot \boldsymbol{\sigma} + \mathbf{b} &= \mathbf{0} \quad \text{in } \Omega_t, \label{eq:momentum_spat}\\
	\boldsymbol{\sigma} \cdot \mathbf{n} &= \bar{\mathbf{t}} \quad \text{on } \Gamma_t, \label{eq:traction_spat}\\
	\frac{G_c}{l_0}\phi - G_c l_0 \nabla_{\mathbf{x}} \cdot (\hat{\mathbf{A}} \cdot \nabla_{\mathbf{x}}\phi) &= -g'(\phi)H \quad \text{in } \Omega_t, \label{eq:phasefield_spat}\\
	\nabla_{\mathbf{x}}\phi \cdot \mathbf{n} &= 0 \quad \text{on } \Gamma_t, \label{eq:phasebc_spat}
\end{align}
where $\boldsymbol{\sigma}$ is the Cauchy stress, $l_0$ is the length scale parameter controlling the width of the diffuse
crack, $\mathbf{b}$ represent the body force vectors in the spatial
configurations, $\mathbf{n}$ is the outward unit normal vectors on the boundary $\Gamma_t$ of the body $\Omega_t$, and $\bar{\mathbf{t}}$ is the traction force. The tensor $\hat{\mathbf{A}}$ is the anisotropic tensor that accounts for the directional dependence of crack resistance due to fiber orientation. This tensor is defined as~\cite{wu2020phase}
\begin{equation}
	\hat{\mathbf{A}} = \mathbf{I} + \hat{\alpha} \mathbf{A}, \label{eq:aniso_tensor}
\end{equation}
where $\mathbf{I}$ is the identity tensor, $\hat{\alpha}$ is a dimensionless parameter controlling the degree of anisotropy. The parameter $\hat{\alpha}$ quantifies how much the fiber orientation influences the crack propagation resistance, with $\hat{\alpha} = 0$ recovering the isotropic phase-field model.

The anisotropic formulation in Eq.~\eqref{eq:phasefield_spat} ensures that cracks preferentially propagate perpendicular to the fiber alignment direction, reflecting the enhanced toughness along fiber directions. This physically represents the fact that crack propagation requires more energy when traversing across fibers compared to propagating parallel to them. The crack resistance is highest in directions where fibers are aligned, as indicated by larger eigenvalues of the orientation tensor.

The energy release rate $G_c$ represents the critical energy required to create a unit area of crack surface. For fiber-reinforced composites, this material parameter can be calibrated from
experimental fracture tests and may depend on the fiber content and orientation.

\subsection{Energy-based crack driving force}
\label{crack_driving_force}

To account for the anisotropic and rate-dependent fracture behavior of fiber-reinforced composites, an energy-based crack driving force is adopted. The crack driving force $\mathcal{H}$ is defined as the maximum strain energy density attained during the loading history by
\begin{equation}
	\mathcal{H}(t) = \max_{\tau \in [0,t]} \mathcal{Y}(\tau),
	\label{eq:history_field}
\end{equation}
where $\mathcal{Y}$ represents the strain energy density that drives crack propagation. The strain energy density is computed from the equilibrium and non-equilibrium contributions to the free energy as
\begin{equation}
	\mathcal{Y} = \psi_{\text{eq}}(\bar{\mathbf{B}}^{\text{ve}}) + \psi_{\text{neq}}(\bar{\mathbf{B}}^e) + \langle\psi_{\text{vol}}(J)\rangle_+,
	\label{eq:energy_density}
\end{equation}
where $\langle \cdot \rangle_+ = (\cdot + |\cdot|)/2$ denotes the positive part operator. The positive part of the volumetric energy is used to distinguish between tension and compression states, preventing crack growth under purely compressive loading.

The volumetric energy contribution is split based on the Jacobian as
\begin{equation}
	\langle\psi_{\text{vol}}(J)\rangle_+ = 
	\begin{cases}
		\psi_{\text{vol}}(J) & \text{if } J \geq 1, \\
		0 & \text{if } J < 1,
	\end{cases}
	\label{eq:vol_split}
\end{equation}
where $J \geq 1$ corresponds to tension (volume expansion) and $J < 1$ corresponds to compression (volume contraction).

This formulation ensures that (1) crack propagation is driven by both elastic (equilibrium) and viscous (non-equilibrium) energy contributions, (2) fracture occurs only under tensile loading conditions, and (3) the maximum energy attained during the loading history is preserved through the history variable $\mathcal{H}$, ensuring irreversibility of crack growth.

The history variable $\mathcal{H}$ in Eq.~\eqref{eq:history_update} ensures the positive evolution of the phase-field variable, preventing crack healing. 
\begin{equation}
	\mathcal{H}^{n+1} = \max(\mathcal{H}^n, \mathcal{Y}^{n+1}),
	\label{eq:history_update}
\end{equation}
where $n$ denotes the time step.

\subsection{Weak form and finite element discretization}

To obtain the weak form of the governing equations, the weighted residual approach is used. Eqs. \eqref{eq:momentum_spat} and \eqref{eq:phasefield_spat} are multiplied by weight functions and integrated over $\Omega_t$. Using the divergence theorem and imposing the boundary conditions, the weak form of the
governing equations can be derived as
\begin{equation}
	\int_{\Omega_t} \boldsymbol{\sigma} : \nabla_{\mathbf{x}}\boldsymbol{\eta}_u \, dv - \int_{\Omega_t} \rho_t \mathbf{b} \cdot \boldsymbol{\eta}_u \, dv - \int_{\Gamma_t} \bar{\mathbf{t}} \cdot \boldsymbol{\eta}_u \, da = 0 \quad \forall \boldsymbol{\eta}_u \in H^1_0(\Omega), \label{eq:weak_u_spat}
\end{equation}
and
\begin{equation}
	\int_{\Omega_t} \left[ J^{-1}g'(\phi)\mathcal{H} \eta_\phi + J^{-1}\frac{G_c}{l_0} \phi \eta_\phi + J^{-1}G_c l_0 \nabla_{\mathbf{x}}\phi \cdot \hat{\mathbf{A}} \cdot \nabla_{\mathbf{x}}\eta_\phi \right] dv = 0 \quad \forall \eta_\phi \in H^1_0(\Omega). \label{eq:weak_phi_spat}
\end{equation}

\subsection{Consistent incremental-iterative scheme}

Assuming that only deformation-independent loads act on the body, Eqs.~\eqref{eq:weak_u_spat} and \eqref{eq:weak_phi_spat} can be expressed in terms of external and internal nodal forces as
\begin{align}
	\mathbf{r}^u &= \mathbf{f}^u_{\text{int}} - \mathbf{f}^u_{\text{ext}} = \mathbf{0}, \label{eq:residual_u}\\
	\mathbf{r}^\phi &= \mathbf{f}^\phi_{\text{int}} - \mathbf{f}^\phi_{\text{ext}} = \mathbf{0}, \label{eq:residual_phi}
\end{align}
where
\begin{align}
	\mathbf{f}^u_{\text{int}} &= \int_{\Omega_t} \boldsymbol{\sigma} : \nabla_x \boldsymbol{\eta}_u \, dv, \label{eq:fint_u}\\
	\mathbf{f}^u_{\text{ext}} &= \int_{\Omega} \rho_t \mathbf{b} \cdot \boldsymbol{\eta}_u \, dv + \int_{\Gamma_t} \bar{\mathbf{t}} \cdot \boldsymbol{\eta}_u \, da, \label{eq:fext_u}\\
	\mathbf{f}^\phi_{\text{int}} &= \int_{\Omega_t} \left[J^{-1}g'(\phi)\mathcal{H}\eta_\phi + J^{-1}\frac{G_c}{l_0}\phi\eta_\phi + J^{-1}G_c l_0 \nabla_x \phi \cdot \hat{\mathbf{A}} \cdot \nabla_x \eta_\phi\right] dv, \label{eq:fint_phi}\\
	\mathbf{f}^\phi_{\text{ext}} &= \mathbf{0}. \label{eq:fext_phi}
\end{align}

By linearizing Eqs.~\eqref{eq:residual_u} and \eqref{eq:residual_phi} at iteration $i + 1$ with respect to the previous iteration $i$, a consistent tangent stiffness is obtained
\begin{align}
	\mathbf{r}^u_{i+1} &= \mathbf{r}^u_i + \Delta \mathbf{r}^u = \mathbf{0}, \label{eq:linearized_u}\\
	\mathbf{r}^\phi_{i+1} &= \mathbf{r}^\phi_i + \Delta \mathbf{r}^\phi = \mathbf{0}, \label{eq:linearized_phi}
\end{align}
where
\begin{align}
	\Delta \mathbf{r}^u &= D_u \mathbf{r}^u_i \cdot \Delta \mathbf{u} + D_\phi \mathbf{r}^u_i \cdot \Delta \phi, \label{eq:delta_ru}\\
	\Delta \mathbf{r}^\phi &= D_u \mathbf{r}^\phi_i \cdot \Delta \mathbf{u} + D_\phi \mathbf{r}^\phi_i \cdot \Delta \phi. \label{eq:delta_rphi}
\end{align}

The linearization of Eqs.~\eqref{eq:weak_u_spat} and \eqref{eq:weak_phi_spat} in spatial formulation yields
\begin{multline}
	\int_{\Omega_t} \left(\nabla_x \Delta \mathbf{u} \cdot \boldsymbol{\sigma} \cdot \nabla_x \boldsymbol{\eta}_u + \nabla^s_x \boldsymbol{\eta}_u : \hat{\mathbf{c}} : \nabla^s_x \Delta \mathbf{u}\right) dv \\
	+ \int_{\Omega_t} \left(\nabla^s_x \boldsymbol{\eta}_u : D_\phi \boldsymbol{\sigma} \cdot \Delta \phi\right) dv = \mathbf{f}^u_{\text{ext}} - \mathbf{f}^u_{\text{int},i},
	\label{eq:linearized_momentum}
\end{multline}
and
\begin{equation}
	\begin{split}
		&\int_{\Omega_t} J^{-1}g'(\phi) 2\frac{\partial \mathcal{H}}{\partial \mathbf{g}} \cdot \nabla_{\mathbf{x}}\Delta\mathbf{u} \, \eta_\phi \, dv \\
		&+ \int_{\Omega_t} \left[ J^{-1}g''(\phi)\mathcal{H}\Delta\phi \, \eta_\phi + J^{-1}\frac{G_c}{l_0}\Delta\phi \, \eta_\phi + J^{-1}G_c l_0 \nabla_{\mathbf{x}}\Delta\phi \cdot \hat{\mathbf{A}} \cdot \nabla_{\mathbf{x}}\eta_\phi \right] dv \\
		&= \mathbf{f}^\phi_{\text{ext}} - \mathbf{f}^\phi_{\text{int},i},
	\end{split}
	\label{eq:linearized_phasefield}
\end{equation}
where $\hat{\mathbf{c}}$ is the spatial tangent modulus defined as the Piola push-forward of the material tangent $\frac{\partial \mathbf{S}}{\partial \mathbf{C}}$, $\mathbf{S}$ is the second Piola-Kirchhoff stress, and $\frac{\partial(\cdot)}{\partial \mathbf{g}} = \mathbf{F}\frac{\partial(\cdot)}{\partial \mathbf{C}}\mathbf{F}^T$.

\subsection{Finite element formulation}

The linearized equilibrium and phase-field equations are assembled into the following system
\begin{equation}
\begin{bmatrix}
	\mathbf{K}^{uu}_i & \mathbf{K}^{u\phi}_i \\
	\mathbf{K}^{\phi u}_i & \mathbf{K}^{\phi\phi}_i
\end{bmatrix}
\begin{bmatrix}
	\Delta \mathbf{u}_{i+1} \\
	\Delta \boldsymbol{\phi}_{i+1}
\end{bmatrix}
=
\begin{bmatrix}
	\mathbf{f}^u_{\text{ext}} \\
	\mathbf{f}^\phi_{\text{ext}}
\end{bmatrix}
-
\begin{bmatrix}
	\mathbf{f}^u_{\text{int},i} \\
	\mathbf{f}^\phi_{\text{int},i}
\end{bmatrix},
\label{eq:fem_system}
\end{equation}
where
\begin{align}
\mathbf{K}^{uu}_i &= \int_{\Omega} \mathbf{B}_u^T \hat{\mathbf{c}} \mathbf{B}_u \, d\Omega + \int_{\Omega} \mathbf{B}_u^T \boldsymbol{\sigma} \mathbf{B}_u \, d\Omega, \label{eq:Kuu}\\
\mathbf{K}^{u\phi}_i &= \int_{\Omega} \mathbf{B}_u^T \left( \frac{\partial \boldsymbol{\sigma}}{\partial \phi} \right) \mathbf{N}_\phi \, d\Omega, \label{eq:Kuphi}\\
\mathbf{K}^{\phi u}_i &= \int_{\Omega} \mathbf{N}_\phi^T J^{-1}g'(\phi) 2\frac{\partial \mathcal{H}}{\partial \mathbf{g}} \mathbf{B}_u \, d\Omega, \label{eq:Kphiu}\\
\mathbf{K}^{\phi\phi}_i &= \int_{\Omega} J^{-1}\left[ \mathbf{N}_\phi^T \left( g''(\phi) \mathcal{H} + \frac{G_c}{l_0} \right) \mathbf{N}_\phi + G_c l_0 \mathbf{B}_\phi^T \hat{\mathbf{A}} \mathbf{B}_\phi \right] d\Omega, \label{eq:Kphiphi}
\end{align}
where $\mathbf{N}_u$ and $\mathbf{N}_\phi$ are the shape function matrices interpolating the nodal values of $\mathbf{u}$ and $\phi$,
respectively, and $\mathbf{B}_u$ and $\mathbf{B}_\phi$ are the gradient operators for the displacement and phase-field,
respectively.

The coupled phase-field fracture problem is solved using a staggered algorithm \cite{miehe2010phase}. In this approach, the displacement and phase-field equations are solved sequentially within each load step.

\subsection{Consistent tangent moduli based on the Jaumann-Zaremba stress rate}

When dealing with large rotations in finite deformation analysis, the material derivative of stress tensors does not transform properly under superposed rigid body motion. To ensure objectivity of the stress rate, the formulation uses the Jaumann-Zaremba rate. The spatial tangent modulus $\hat{\mathbf{c}}$, required
by the FE weak form in Eq.~\eqref{eq:Kuu}, is defined as the Piola
push-forward of the material tangent $\partial\mathbf{S}/\partial\mathbf{C}$,
\begin{equation}
	\hat{\mathbf{c}} = \frac{1}{J}\,
	\left(\mathbf{F}\,\overline{\otimes}\,\mathbf{F}\right)
	: \frac{\partial\mathbf{S}}{\partial\mathbf{C}} :
	\left(\mathbf{F}^T\,\overline{\otimes}\,\mathbf{F}^T\right).
	\label{eq:c_hat_pushforward}
\end{equation}

A closed-form evaluation of Eq.~\eqref{eq:c_hat_pushforward} is not
straightforward for the complex viscoelastic-viscoplastic-damage model.
Instead, $\hat{\mathbf{c}}$ is obtained in two steps: first, the Jaumann tangent $\mathbf{C}^{\sigma J}$ is computed numerically via finite differences~\cite{sun2008numerical}; second, $\hat{\mathbf{c}}$ is recovered analytically from $\mathbf{C}^{\sigma J}$ using the correction derived below.

The Jaumann rate of the Cauchy stress is expressed as
\begin{equation}
	\overset{\nabla}{\boldsymbol{\sigma}}
	= \dot{\boldsymbol{\sigma}}
	- \mathbf{W}\boldsymbol{\sigma}
	- \boldsymbol{\sigma}\mathbf{W}^T
	= \mathbf{C}^{\sigma J} : \mathbf{D},
	\label{eq:jaumann_rate}
\end{equation}
where $\mathbf{D}$ is the rate of deformation tensor and $\mathbf{W}$ is the spin tensor. To numerically compute $\mathbf{C}^{\sigma J}$, small perturbations are applied to components of the deformation gradient. For a perturbation in the
$(i,j)$ component,
\begin{equation}
	\Delta\mathbf{F}_{ij} = \frac{\epsilon}{2}
	\bigl(\mathbf{e}_i \otimes \mathbf{e}_j\,\mathbf{F}
	+ \mathbf{e}_j \otimes \mathbf{e}_i\,\mathbf{F}\bigr),
	\label{eq:perturbation}
\end{equation}
where $\epsilon = 10^{-5}$ is the perturbation parameter and $\mathbf{e}_i$
are Cartesian basis vectors. The corresponding perturbations in the spin and
rate-of-deformation tensors are
\begin{align}
	\Delta\mathbf{W}_{ij} &= \mathbf{0},
	\label{eq:delta_W}\\
	\Delta\mathbf{D}_{ij} &= \frac{\epsilon}{2}
	\bigl(\mathbf{e}_i \otimes \mathbf{e}_j
	+ \mathbf{e}_j \otimes \mathbf{e}_i\bigr).
	\label{eq:delta_D}
\end{align}
The perturbed deformation gradient and the resulting stress increment are
\begin{align}
	\hat{\mathbf{F}}_{ij} &= \mathbf{F} + \Delta\mathbf{F}_{ij},
	\label{eq:F_perturbed}\\
	\Delta\boldsymbol{\sigma}
	&\approx \boldsymbol{\sigma}\!\left(\hat{\mathbf{F}}_{ij}\right)
	- \boldsymbol{\sigma}(\mathbf{F}).
	\label{eq:delta_sigma}
\end{align}
Since $\Delta\mathbf{W}_{ij} = \mathbf{0}$, the linearized incremental form of
Eq.~\eqref{eq:jaumann_rate} reduces to $\Delta\boldsymbol{\sigma} =
\mathbf{C}^{\sigma J}:\Delta\mathbf{D}$, so the components of
$\mathbf{C}^{\sigma J}$ follow directly as
\begin{equation}
	\mathbf{C}^{\sigma J}_{ij}
	= \frac{1}{\epsilon}
	\Bigl[\boldsymbol{\sigma}\!\left(\hat{\mathbf{F}}_{ij}\right)
	- \boldsymbol{\sigma}(\mathbf{F})\Bigr].
	\label{eq:CsigJ_numerical}
\end{equation}

Algorithm~\ref{alg:tangent_algorithm} summarizes the step-by-step method for computing the consistent tangent modulus required for the FE analysis.

\begin{algorithm}[H]
\centering
\caption{Algorithm for computing the consistent tangent modulus.}
\label{alg:tangent_algorithm}
\begin{tabular}{l}
	\textbf{1.} Define perturbation parameter: $\epsilon = 10^{-5}$ \\
	\textbf{2.} Compute the right Cauchy-Green tensor: $\mathbf{C} = \mathbf{F}^T\mathbf{F}$ \\
	\textbf{3.} Calculate the Cauchy stress $\boldsymbol{\sigma}$ using equations from Section 2 \\
	\textbf{4.} \textbf{for} $k = 1, \ldots, 3$ \textbf{do} \\
	\textbf{5.} \quad \textbf{for} $l = 1, \ldots, 3$ \textbf{do} \\
	\textbf{6.} \quad \quad Initialize $\hat{\mathbf{F}} = \mathbf{F}$ \\
	\textbf{7.} \quad \quad Perturb: $\hat{F}_{kl} = \hat{F}_{kl} + \epsilon/2$ \\
	\textbf{8.} \quad \quad Perturb: $\hat{F}_{lk} = \hat{F}_{lk} + \epsilon/2$ \\
	\textbf{9.} \quad \quad Calculate perturbed stress response: $\Delta \boldsymbol{\sigma}$ using Eq.~\eqref{eq:delta_sigma} \\
	\textbf{10.} \quad \quad \textbf{for} $i = 1, \ldots, 3$ \textbf{do} \\
	\textbf{11.} \quad \quad \quad \textbf{for} $j = 1, \ldots, 3$ \textbf{do} \\
	\textbf{12.} \quad \quad \quad \quad Compute and store $C^{\sigma J}_{ijkl}$ and $C^{\sigma J}_{ijlk}$ using Eq.~\eqref{eq:CsigJ_numerical} \\
	\textbf{13.} \quad \quad \quad \textbf{end} \\
	\textbf{14.} \quad \quad \textbf{end} \\
	\textbf{15.} \quad \textbf{end} \\
	\textbf{16.} \textbf{end} \\
	\textbf{17.} Determine the material tangent: $\mathbf{C}^{\sigma J}$ \\
	\textbf{18.} Store the tangent tensor in Voigt notation \\
	\textbf{19.} Solve the system of equations using a staggered algorithm \\
\end{tabular}
\end{algorithm}

The Jaumann tangent $\mathbf{C}^{\sigma J}$ and the spatial tangent
$\hat{\mathbf{c}}$ are conjugate to the Jaumann rate and the Truesdell (Lie) rate, respectively. Recovering $\hat{\mathbf{c}}$ from $\mathbf{C}^{\sigma J}$ gives

\begin{equation}
	\hat{\mathbf{c}} = \mathbf{C}^{\sigma J}
	- \tfrac{1}{2}\bigl(
	\mathbf{I}\,\overline{\otimes}\,\boldsymbol{\sigma}
	+ \mathbf{I}\,\underline{\otimes}\,\boldsymbol{\sigma}
	+ \boldsymbol{\sigma}\,\overline{\otimes}\,\mathbf{I}
	+ \boldsymbol{\sigma}\,\underline{\otimes}\,\mathbf{I}
	\bigr)
	+ \boldsymbol{\sigma} \otimes \mathbf{I}.
	\label{eq:chat_from_CsigJ}
\end{equation}

The numerical approximation of the tangent modulus is computationally efficient and accurate for the complex constitutive behavior of fiber-reinforced composites with coupled viscoelasticity, viscoplasticity, and damage.

\section{Numerical simulations}
\label{sec:results}

A comprehensive set of numerical simulations is presented to demonstrate the predictive capability of the proposed framework and to reveal the complex interplay between viscoelastic-viscoplastic constitutive behavior, anisotropic crack propagation, and hygrothermal environmental effects in SFRPs. All simulations examine GF-reinforced epoxy composites under quasi-static loading. The material parameters for the viscoelastic-viscoplastic constitutive model and phase-field fracture formulation are summarized in Table~\ref{tab:param}, with values calibrated from experimental data reported in the literature for the same composite material. The numerical investigations explore the influence of fiber orientation, volume fraction, environmental conditions (moisture and temperature), fiber mechanical properties, and laminate architecture on both the global mechanical response and crack path evolution. In the following simulations, the saturated moisture condition refers to 1 wt\% moisture content.

\begin{table}[h]
\centering
\caption{Material parameters used in the numerical simulations.} \label{tab:param}
\begin{footnotesize}
	\begin{tabular}{lcccc}
		\toprule 
		Parameter & Symbol & Value & Equation & References \\
		\midrule 
		Equilibrium shear modulus & $\mu_{\text{eq}}^{0}$ (MPa) & 760 & (29) & \cite{BAHTIRI2023116293}  \\
		Non-equilibrium shear modulus  & $\mu_{\text{neq}}^{0}$ (MPa) & 790 & (29) & \cite{BAHTIRI2023116293} \\
		Volumetric bulk modulus & $k_{v}^{0}$ (MPa) & 1154 & (38) & \cite{BAHTIRI2023116293} \\
		Viscoelastic pre-factor & $\dot{\varepsilon}_{0}$ ($\text{s}^{-1}$) & $1.0447 \times 10^{12}$ & (13) & \cite{BAHTIRI2023116293} \\
		Activation energy & $\Delta H$ (J) & $1.977 \times 10^{-19}$ & (13) & \cite{BAHTIRI2023116293}  \\
		Stress exponent & $m$ & 0.657 & (13) & \cite{BAHTIRI2023116293}\\
		Athermal yield stress & $\tau_{0}$ (MPa) & 40 & (13) & \cite{BAHTIRI2023116293} \\
		Viscoplastic parameter & $a$ & 22$\omega_w$ + 0.8 & (18) & -- \\
		Viscoplastic exponent & $b$ & 1.1 & (18) & -- \\
		Viscoplastic threshold & $\sigma_{0}$ (MPa) & 30 & (18) & -- \\
		Critical energy release rate & $G_{c}$ (N/mm) & 0.19 & (55) & \cite{arash2023effect} \\
		Length scale parameter & $l_{0}$ (mm) & 0.02 & (55) & \cite{arash2023effect} \\	
		Moisture swelling coefficient & $\alpha_{w}$ & 0.039 & (4) & \cite{arash2023effect} \\
		Thermal expansion coefficient & $\alpha_{\theta}$ ($\text{K}^{-1}$) & $4.19 \times 10^{-5}$ & (3) & \cite{arash2023effect} \\
		Temperature sensitivity & $\alpha$ ($\text{K}^{-1}$) & 0.01093 & (30)-(31) & \cite{BAHTIRI2023116293} \\
		Fiber stiffness parameter & $a_1$ & 9 & (35) & -- \\
		Fiber stiffness parameter & $a_2$ & 1 & (35) & -- \\
		Fiber stiffness parameter & $a_3$ & 1 & (35) & -- \\
		\bottomrule
	\end{tabular}
\end{footnotesize}
\end{table}

\begin{figure}[H]
\centering
\begin{subfigure}[b]{0.45\linewidth}
	\centering
	\includegraphics{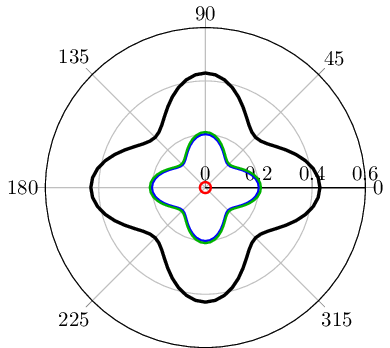}
	\caption{ }
	\label{fig:driving_force0_1}
\end{subfigure}\hspace{24pt}
\begin{subfigure}[b]{0.45\linewidth}
	\centering
	\includegraphics{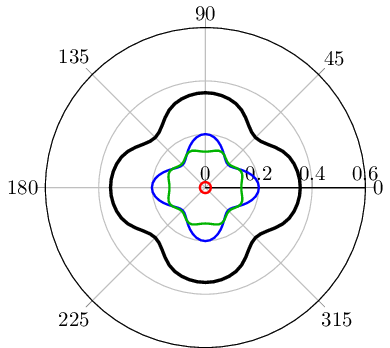}
	\caption{ }
	\label{fig:driving_force0_2}
\end{subfigure}

\begin{subfigure}[b]{0.45\linewidth}
	\centering
	\includegraphics{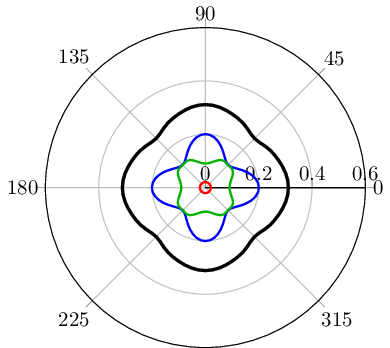}
	\caption{ }
	\label{fig:driving_force0_3}
\end{subfigure}\hspace{24pt}
\begin{subfigure}[b]{0.45\linewidth}
	\centering
	\includegraphics{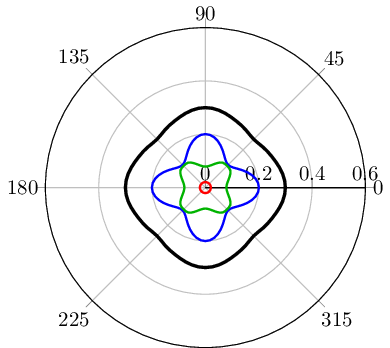}
	\caption{ }
	\label{fig:driving_force0_4}
\end{subfigure}
\caption{Spatial distribution of crack driving force components during viscous relaxation for GF (50 wt\%)/epoxy with balanced fiber families oriented at $\pm 45^{\circ}$ under moisture-saturated conditions at 300~K. The contours show (a) initial response at $t = 10^{-6}$~s, (b) early relaxation at $t = 0.005$~s, (c) intermediate relaxation at $t = 0.05$~s, and (d) near-equilibrium at $t = 0.1$~s. Color coding: total (black), equilibrium (blue), non-equilibrium (green), and volumetric (red) energy contributions.}
\label{fig:driving_force0}
\end{figure}

\begin{figure}[H]
\centering
\begin{subfigure}[b]{0.45\linewidth}
	\centering
	\includegraphics{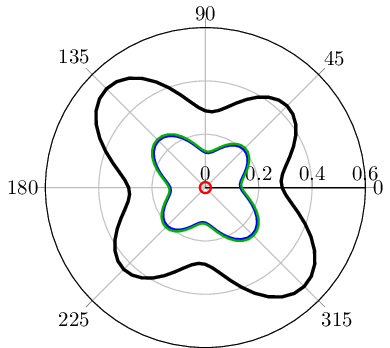}
	\caption{ }
	\label{fig:driving_force1_1}
\end{subfigure}\hspace{24pt}
\begin{subfigure}[b]{0.45\linewidth}
	\centering
	\includegraphics{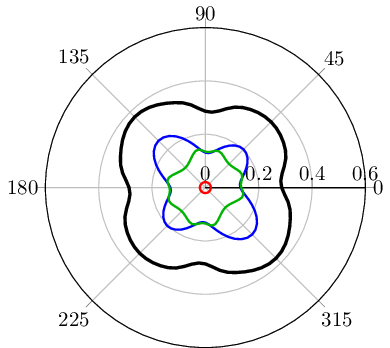}
	\caption{ }
	\label{fig:driving_force1_2}
\end{subfigure}

\begin{subfigure}[b]{0.45\linewidth}
	\centering
	\includegraphics{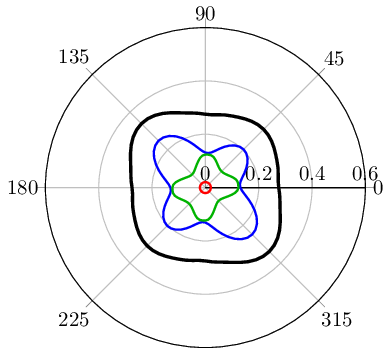}
	\caption{ }
	\label{fig:driving_force1_3}
\end{subfigure}\hspace{24pt}
\begin{subfigure}[b]{0.45\linewidth}
	\centering
	\includegraphics{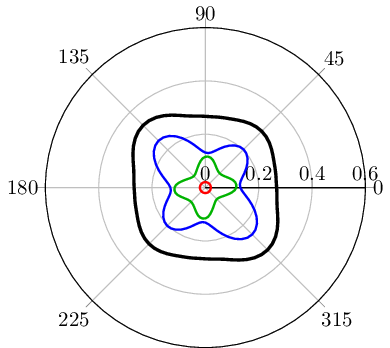}
	\caption{ }
	\label{fig:driving_force1_4}
\end{subfigure}
\caption{Spatial distribution of crack driving force components during viscous relaxation for moisture-saturated GF (50 wt\%)/epoxy with unbalanced fiber distribution (70 wt\% at $-45^{\circ}$ and 30 wt\% at $+45^{\circ}$) at 300~K. The contours show (a) initial response at $t = 10^{-6}$~s, (b) early relaxation at $t = 0.005$~s, (c) intermediate relaxation at $t = 0.05$~s, and (d) near-equilibrium at $t = 0.1$~s. Color coding: total (black), equilibrium (blue), non-equilibrium (green), and volumetric (red) energy contributions.}
\label{fig:driving_force}
\end{figure}

We first explore how viscoelastic and viscoplastic energy storage mechanisms contribute differently to the crack driving force in the composites. Figs.~\ref{fig:driving_force0} and~\ref{fig:driving_force} present polar plots of the crack driving force $\mathcal{H}$ at four time instants following the application of an instantaneous tensile load for GF (50 wt\%)/epoxy composites with different fiber distributions. Fig.~\ref{fig:driving_force0} examines a balanced distribution with 50 wt\% aligned at $45^{\circ}$ and 50 wt\% at $-45^{\circ}$, while Fig.~\ref{fig:driving_force} investigates an unbalanced configuration with 30 wt\% at $45^{\circ}$ and 70 wt\% at $-45^{\circ}$. An initial tensile load is applied by prescribing an initial deformation gradient, after which the crack driving force is recorded as the system undergoes viscous relaxation at four time instants--$t = 10^{-6}$~s, $0.005$~s, $0.05$~s, and $0.1$~s--representing the initial elastic response, early viscous relaxation, intermediate relaxed state, and near-equilibrium state, respectively. The total crack driving force (black curve), equilibrium (blue curve), non-equilibrium (green curve), and volumetric (red curve) energy components are shown separately to reveal their individual contributions.

For the balanced fiber distribution (Fig.~\ref{fig:driving_force0}), the polar representation reveals a four-lobed pattern with maximum energy accumulation occurring along $0^{\circ}$ and $90^{\circ}$ directions rather than along the fiber orientations. This behavior arises because the symmetric $\pm 45^{\circ}$ fiber arrangement creates a quasi-isotropic material response. At the initial stage ($t = 10^{-6}$~s), the total crack driving force reaches approximately 0.5 MPa along these directions, decreasing to approximately 0.3 MPa at quasi-equilibrium ($t = 0.1$~s)--a 40\% reduction due to viscous dissipation. The equilibrium energy (blue curve) remains relatively constant throughout relaxation, while the non-equilibrium energy (green curve) shows dramatic reduction as viscous mechanisms redistribute stored energy, and the volumetric component (red curve) increases monotonically to become the dominant contribution at later stages. In contrast, the unbalanced distribution (Fig.~\ref{fig:driving_force}) produces an asymmetric pattern reflecting the directional bias introduced by the 70-30 fiber distribution. The crack driving force exhibits higher magnitude lobes in directions influenced by the dominant $-45^{\circ}$ fiber family, demonstrating how fiber distribution asymmetry breaks the quasi-isotropic behavior and creates preferential directions for crack propagation. The temporal evolution follows similar trends as the balanced case, with viscous relaxation causing approximately 40\% reduction in total driving force over the same time scale, but the directional anisotropy persists throughout the loading history, confirming that fiber architecture fundamentally controls both the magnitude and spatial distribution of energy available to drive fracture.

\begin{figure}[H]
\centering
	\scalebox{0.8}{%
		\includegraphics{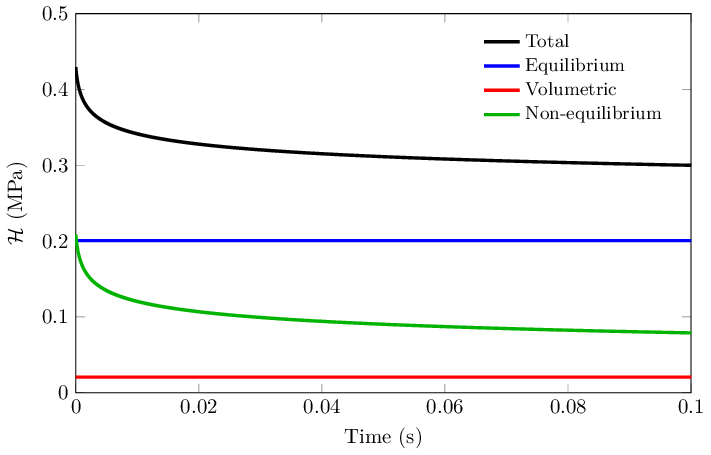}
	}
\caption{Temporal evolution of crack driving force components for moisture-saturated GF (50 wt\%)/epoxy with unbalanced fiber distribution (70 wt\% at $-45^{\circ}$ and 30 wt\% at $+45^{\circ}$) at 300~K under uniaxial tensile loading applied at $45^{\circ}$ to the horizontal axis.}
\label{fig:energy_evolution}
\end{figure}

The temporal decomposition of energy contributions provides further insight into the rate-dependent fracture mechanisms. Fig.~\ref{fig:energy_evolution} shows the evolution of total crack driving force $\mathcal{H}$ and its constituent components--equilibrium, non-equilibrium (viscous), and volumetric energies--for the unbalanced fiber distribution. The initial tensile load is applied along $45^{\circ}$. The total driving force exhibits rapid initial increase followed by more gradual evolution as time-dependent viscous processes activate. The equilibrium energy remains constant after initial loading, reflecting the elastic storage capacity, while the non-equilibrium energy decreases from an initial peak as viscous relaxation occurs through chain sliding and fiber-matrix interfacial effects. The volumetric energy contribution increases monotonically and becomes the dominant component at later loading stages. This decomposition reveals a fundamental characteristic of fracture in viscoelastic-viscoplastic composites that cannot be captured by elastic phase-field models: the crack driving force comprises time-varying contributions from multiple physical mechanisms, with viscous energy dominating at early times (reflecting rate-dependent resistance) and volumetric energy becoming predominant near failure (indicating transition toward tensile fracture). The ability to track these individual contributions provides unprecedented insight into the relative importance of different deformation mechanisms throughout the loading history and enables identification of which physical processes control fracture initiation versus propagation.

\begin{figure}[H]
\centering
\includegraphics{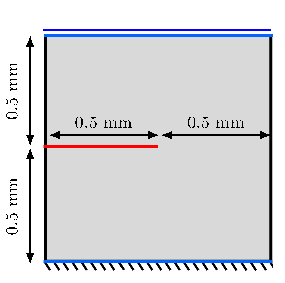}
\caption{Geometry and boundary conditions of single edge notched specimens.}
\label{fig:sn_contour}	
\end{figure}

The proposed phase-field model's ability to predict fracture patterns is evaluated by conducting single-edge notched tensile tests. The geometry and boundary conditions of the tests are shown in Figure~\ref{fig:sn_contour}. The tensile load is applied at a deformation rate of $\dot{u}=1$~mm/min with constant displacement increments of $10^{-6}$~mm. 

\begin{figure}[H]
\centering
\begin{subfigure}[b]{0.5\linewidth}
	\centering
	\includegraphics{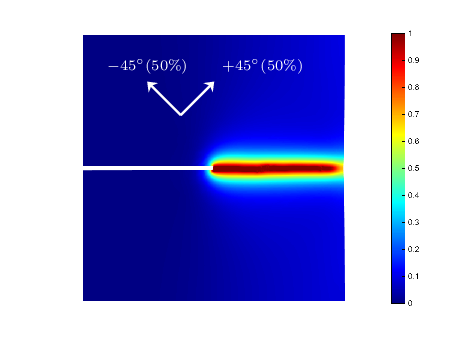}
	\caption{ }
	\label{fig:fracture_family_0}
\end{subfigure}%
\begin{subfigure}[b]{0.5\linewidth}
	\centering	
	\includegraphics{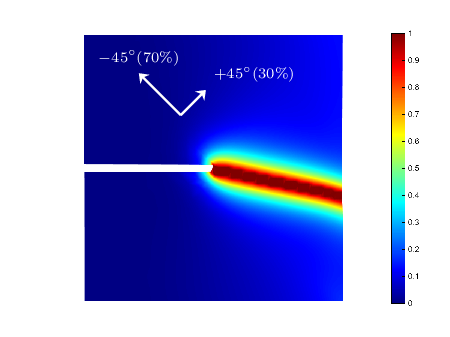}
	\caption{}			
	\label{fig:fracture_family_1}
\end{subfigure}	
\caption{Effect of fiber family members on the crack path in moisture-saturated GF(50 wt\%)/epoxy at 300~K: (a) balanced fiber distribution (50-50 at $\mp 45^{\circ}$), and (b) unbalanced distribution (70 wt\% at $-45^{\circ}$ and 30 wt\% at $+45^{\circ}$).}
\label{fig:damagecontour_family}
\end{figure}

\begin{figure}[H]
\centering
\includegraphics{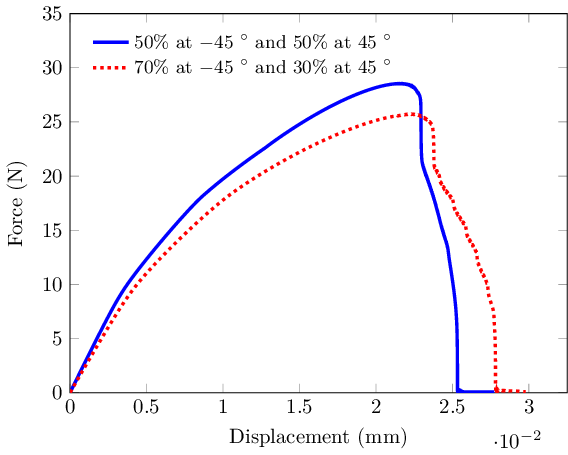}
\caption{Effect of fiber family members on the force--displacement response of moisture-saturated GF(50 wt\%)/epoxy at $T=300$~K.}
\label{fig:forcedisp_family}
\end{figure}

The influence of fiber family distribution on fracture behavior is examined in Figs.~\ref{fig:damagecontour_family} and~\ref{fig:forcedisp_family} for saturated GF(50 wt\%)/epoxy composites at 300 K, comparing two distinct bimodal fiber configurations with the aisotropic parameter $\hat{\alpha}=10$. Fig.~\ref{fig:fracture_family_0} shows a balanced distribution with 50 wt\% fibers aligned at $-45^{\circ}$ and 50 wt\% at $45^{\circ}$, representing a symmetric reinforcement architecture. In contrast, Fig.~\ref{fig:fracture_family_1} presents an unbalanced distribution with 70 wt\% fibers at $-45^{\circ}$ and only 30 wt\% at $45^{\circ}$, creating directional asymmetry in the composite's mechanical response. These configurations enable investigation of how the relative proportions of fiber families--rather than total fiber content or orientation angles themselves--affect crack path evolution and load-bearing capacity.

The crack path visualizations reveal pronounced sensitivity to fiber family distribution. For the balanced 50-50 configuration (Fig.~\ref{fig:fracture_family_0}), the crack propagates in a straight horizontal path perpendicular to the loading direction. This behavior reflects the symmetric reinforcement where neither fiber family dominates the crack deflection tendency, resulting in a compromise trajectory that minimizes the total energy required for propagation through the competing orientations. In contrast, the unbalanced 70-30 distribution (Fig.~\ref{fig:fracture_family_1}) produces substantial crack deflection toward the direction perpendicular to the dominant fiber family ($-45^{\circ}$). The crack path curves noticeably, deviating by approximately $20^{\circ}$ from the horizontal as it preferentially propagates through regions of lower fracture resistance. This deflection occurs because the higher concentration of fibers at $-45^{\circ}$ creates a stronger resistance to crack propagation perpendicular to this orientation, forcing the crack to seek paths that minimize intersection with the dominant fiber family.

The mechanical consequences of fiber family distribution are quantified in Fig.~\ref{fig:forcedisp_family}, which compares force-displacement responses for both configurations. The balanced 50-50 distribution achieves a peak load of approximately 28 N at 0.022 mm displacement, representing superior load-carrying capacity. In contrast, the unbalanced 70-30 configuration reaches only about 26 N peak load at 0.024 mm displacement--roughly a 7\% reduction in strength accompanied by an increase in ductility. Both curves exhibit similar initial stiffness, confirming that elastic properties are primarily determined by total fiber content rather than distribution. However, the onset of damage and subsequent softening behavior differ significantly. The balanced distribution demonstrates more abrupt failure after peak load, with a steep drop indicating rapid crack propagation once the critical threshold is exceeded. This brittle post-peak response reflects the difficulty of crack deflection in a symmetric architecture where both fiber families provide comparable resistance. Conversely, the unbalanced distribution exhibits more gradual post-peak softening with an extended tail, suggesting stable crack growth through progressive damage accumulation.

\begin{figure}[H]
\centering
\begin{subfigure}[b]{0.5\linewidth}
	\centering
	\includegraphics{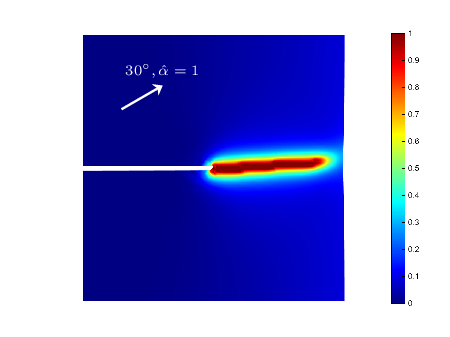}
	\caption{ }
	\label{fig:fracture_30-degree_w1_alpha1}
\end{subfigure}
\begin{subfigure}[b]{0.5\linewidth}
	\centering	
	\includegraphics{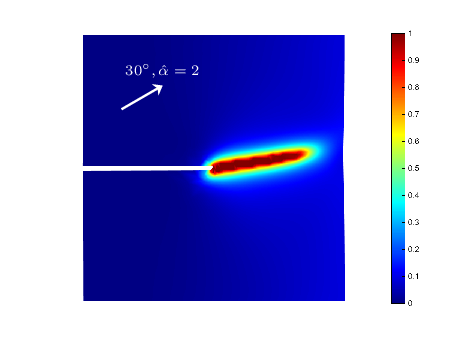}
	\caption{}			
	\label{fig:fracture_30-degree_w1_alpha2}
\end{subfigure}	
\begin{subfigure}[b]{0.5\linewidth}
	\centering	
	\includegraphics{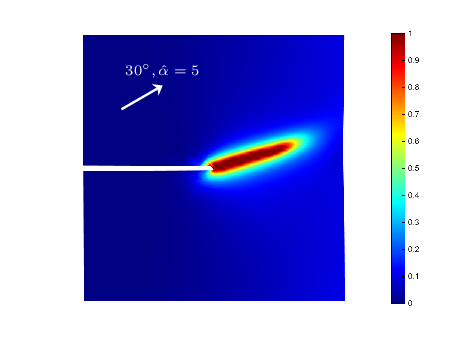}
	\caption{}			
	\label{fig:fracture_30-degree_w1_alpha5}
\end{subfigure}%
\begin{subfigure}[b]{0.5\linewidth}
	\centering	
	\includegraphics{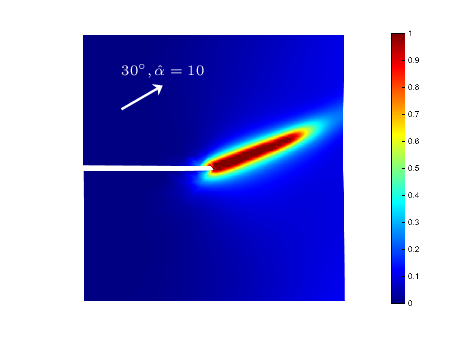}
	\caption{}			
	\label{fig:fracture_30-degree_w1_alpha10}
\end{subfigure}
\caption{Effect of $\hat{\alpha}$ on the crack path in moisture-saturated GF(50 wt\%)/epoxy with fibers aligned at $30^\circ$ at 300~K: (a) $\hat{\alpha}=1$, (b) $\hat{\alpha}=2$, (c) $\hat{\alpha}=5$, and (d) $\hat{\alpha}=10$.}
\label{fig:damagecontour_30-degree}
\end{figure}

The quantitative relationships between material anisotropy and crack path deflection through the anisotropic phase-field formulation is explored in Figs~\ref{fig:fracture_30-degree_w1_alpha1}--\ref{fig:fracture_30-degree_w1_alpha10}. The anisotropy parameter $\hat{\alpha}$ in Eq.~\eqref{eq:aniso_tensor} controls the degree to which fiber orientation influences crack propagation resistance, providing a direct link between microstructural architecture and macroscopic fracture behavior. Fig.~\ref{fig:damagecontour_30-degree} illustrates this relationship for GF(50 wt\%)/epoxy composites with fibers predominantly aligned at $30^{\circ}$ to the horizontal direction, examining four values, $\hat{\alpha} = 1, 2, 5, \text{and} 10$. For the lowest anisotropy, the crack propagates in a relatively straight path perpendicular to loading, exhibiting only minor deflection. As $\hat{\alpha}$ increases to 2, more pronounced deflection toward the fiber direction emerges, resulting in crack path deviation of approximately 10-15° from horizontal. At $\hat{\alpha} = 5$, substantial deflection occurs with the path curving to align more closely perpendicular to fibers. For the highest anisotropy ($\hat{\alpha} = 10$), maximum deflection closely follows the direction of weakest resistance imposed by the fiber architecture. The progressive increase in deflection angle with $\hat{\alpha}$ confirms the physical interpretation of this parameter as controlling the strength of fiber-induced directional toughening, and the systematic nature of this relationship provides quantitative guidance for calibrating $\hat{\alpha}$ from experimental fracture tests on composites with known fiber orientations.

\begin{figure}[H]
\centering
\includegraphics{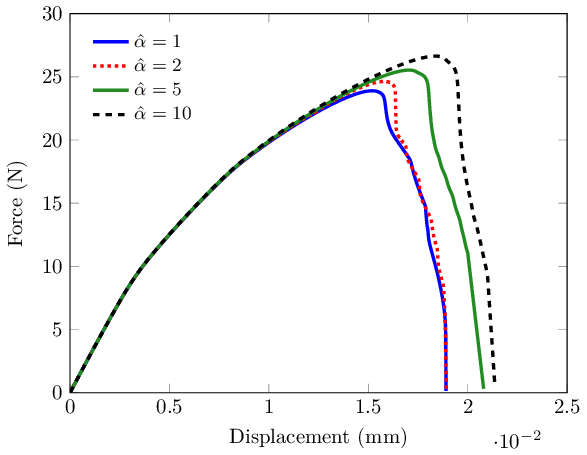}
\caption{Effect of $\hat{\alpha}$ on the force--displacement response of moisture-saturated GF(50 wt\%)/epoxy with fibers aligned at $30^\circ$ at $T=300$~K.}
\label{fig:forcedisp_30-degree}
\end{figure}

The mechanical consequences of crack path deflection are quantified in the corresponding force-displacement responses in Fig.~\ref{fig:forcedisp_30-degree}. All curves exhibit similar initial stiffness since the anisotropy parameter affects crack propagation resistance rather than elastic properties. However, both peak load and displacement at failure increase systematically with $\hat{\alpha}$: from approximately 24~N at 0.016~mm for $\hat{\alpha} = 1$, to 25~N for $\hat{\alpha} = 2$, 26~N for $\hat{\alpha} = 5$, and 27~N at 0.017~mm for $\hat{\alpha} = 10$. The monotonic increase in area under each curve—representing total fracture energy—with $\hat{\alpha}$ confirms that stronger anisotropy enhances composite toughness by forcing cracks along energetically unfavorable paths. 

\begin{figure}[H]
\centering
\begin{subfigure}[b]{0.5\linewidth}
	\centering
	\includegraphics{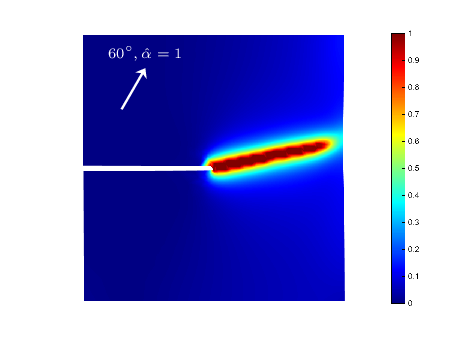}
	\caption{ }
	\label{fig:fracture_60-degree_w1_alpha1}
\end{subfigure}
\begin{subfigure}[b]{0.5\linewidth}
	\centering	
	\includegraphics{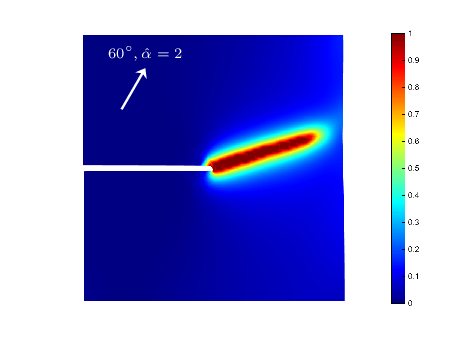}
	\caption{}			
	\label{fig:fracture_60-degree_w1_alpha2}
\end{subfigure}	
\begin{subfigure}[b]{0.5\linewidth}
	\centering	
	\includegraphics{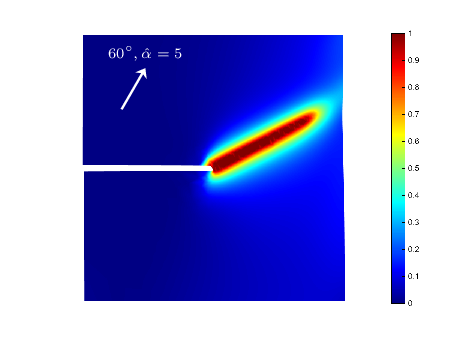}
	\caption{}			
	\label{fig:fracture_60-degree_w1_alpha5}
\end{subfigure}%
\begin{subfigure}[b]{0.5\linewidth}
	\centering	
	\includegraphics{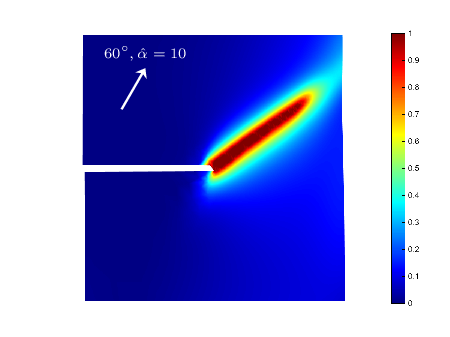}
	\caption{}			
	\label{fig:fracture_60-degree_w1_alpha10}
\end{subfigure}
\caption{Effect of the anisotropy parameter ($\hat{\alpha}$) on the crack path in moisture-saturated GF(50 wt\%)/epoxy with fibers aligned at $60^\circ$ at 300~K: (a) $\hat{\alpha}=1$, (b) $\hat{\alpha}=2$, (c) $\hat{\alpha}=5$, and (d) $\hat{\alpha}=10$.}
\label{fig:damagecontour_60-degree}
\end{figure}

\begin{figure}[H]
\centering
\includegraphics{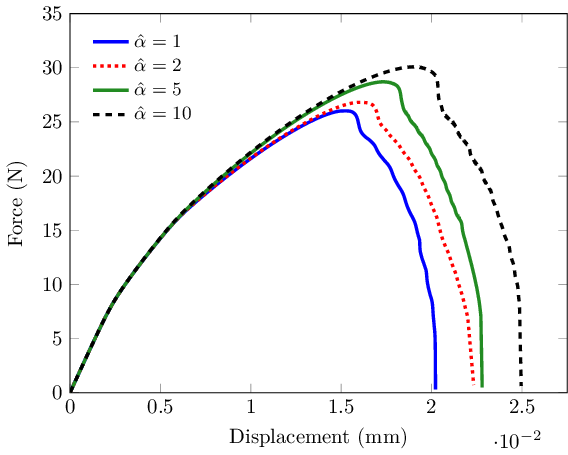}
\caption{Effect of the anisotropy parameter ($\hat{\alpha}$) on the force--displacement response of moisture-saturated GF(50 wt\%)/epoxy with fibers aligned at $60^\circ$ at $T=300$~K.}
\label{fig:forcedisp_60-degree}
\end{figure}

To further demonstrate the generality of the anisotropic phase-field formulation across different fiber orientations, Figs.~\ref{fig:damagecontour_60-degree} and~\ref{fig:forcedisp_60-degree} examine GF(50 wt\%)/epoxy composites with fibers predominantly aligned at $60^{\circ}$ to the horizontal direction under saturated conditions at 300 K. Fig.~\ref{fig:damagecontour_60-degree} illustrates crack path evolution for four anisotropy parameters: $\hat{\alpha} = 1, 2, 5,$ and $10$. For the lowest anisotropy ($\hat{\alpha} = 1$), the crack maintains an approximately horizontal trajectory perpendicular to loading, showing minimal deflection despite the steep fiber orientation. As $\hat{\alpha}$ increases to 2, noticeable deflection emerges with the crack path deviating approximately $20{-}25^{\circ}$ from horizontal--more pronounced than the $30^{\circ}$ fiber orientation case at the same anisotropy parameter, reflecting stronger geometric incompatibility. At $\hat{\alpha} = 5$, substantial deflection manifests with the crack curving significantly to propagate nearly perpendicular to the fiber orientation, exhibiting pronounced waviness and broader damage zones. For maximum anisotropy ($\hat{\alpha} = 10$), extreme deflection occurs, demonstrating the phase-field model's capability to capture spontaneous path selection driven purely by anisotropic material properties.

The mechanical consequences quantified in Fig.~\ref{fig:forcedisp_60-degree} reveal systematic trends consistent with the $30^{\circ}$ orientation results but with quantitative differences reflecting the steeper fiber angle. Peak loads increase progressively with $\hat{\alpha}$ from approximately 25 N at 0.016 mm for $\hat{\alpha} = 1$, to 27 N at 0.017 mm for $\hat{\alpha} = 2$, reaching 30 N at 0.018 mm for $\hat{\alpha} = 5$, and achieving 32 N at 0.019 mm for $\hat{\alpha} = 10$. These peak loads are higher than observed for the $30^{\circ}$ orientation at equivalent anisotropy levels, reflecting the geometric advantage where $60^{\circ}$ fiber alignment provides more effective load transfer through direct fiber stretching.

\begin{figure}[H]
\centering
\includegraphics{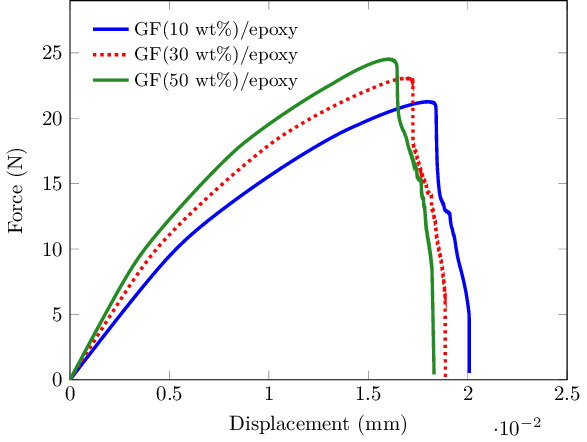}
\caption{Effect of fiber content on the force--displacement response of randomly distributed GF/epoxy at $T=300$~K and moisture-saturated condition.}
\label{fig:forcedisp_rand-degree}
\end{figure}

In the following, the multiphysics coupling between fiber reinforcement, hygrothermal environment, and fracture behavior is investigated. Fig.~\ref{fig:forcedisp_rand-degree} examines the effect of fiber content on the force-displacement response for randomly distributed short fiber/epoxy composites. GF(10 wt\%)/epoxy, GF(30 wt\%)/epoxy, and GF(50 wt\%)/epoxy composites are under saturated condition at $T = 300$~K. The force-displacement curves reveal enhancement of both stiffness and peak load with increasing fiber content, from approximately 21~N at 0.019~mm for 10 wt\%, to 23~N at 0.018~mm for 30 wt\%, and 25~N at 0.017~mm for 50 wt\%. Notably, displacement at peak load decreases with fiber content, indicating that while higher fiber loading enhances load-carrying capacity, it also promotes more brittle fracture with reduced ductility. This trend reflects competing effects: short fibers increase stiffness and strength through load transfer to the stiffer reinforcing phase, but also introduce stress concentration sites at fiber ends and interfaces that promote earlier crack initiation. Nevertheless, fracture energy (area under curves) increases with fiber content despite reduced ductility, confirming the net toughening effect of short fiber reinforcement and demonstrating the model's capability to capture stiffness-toughness-ductility trade-offs inherent in particulate reinforcement.

\begin{figure}[H]
\centering
\includegraphics{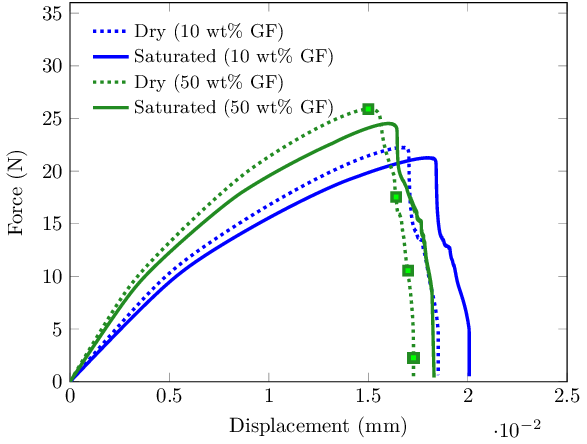}
\caption{Effect of moisture content on the force--displacement response of randomly distributed GF/epoxy at $T=300$~K.}
\label{fig:forcedisp_rand-degree_moisture}
\end{figure}

Moisture absorption effects are quantified in Fig.~\ref{fig:forcedisp_rand-degree_moisture}, comparing dry and saturated conditions for both GF(10 wt\%)/epoxy and GF(50 wt\%)/epoxy at 300~K. For the 10 wt\% system, moisture reduces peak force from approximately 26~N (dry) to 21~N (saturated)—a 19\% degradation—while shifting failure to slightly higher displacement (0.019~mm vs. 0.017~mm), indicating increased matrix ductility through plasticization. For the 50 wt\% composite, moisture causes smaller relative degradation (27~N dry to 25~N saturated, approximately 7\% reduction), suggesting that fiber reinforcement provides some moisture resistance. However, the absolute force reduction is comparable for both compositions (5-6~N), indicating moisture primarily affects matrix-dominated properties. The post-peak behavior changes significantly: saturated specimens exhibit more gradual softening with extended tails, while dry specimens show sharper load drops, reflecting competing effects of matrix plasticization (increasing ductility) and fiber-matrix interface degradation (promoting debonding and distributed damage). The model captures these moisture effects through temperature- and moisture-dependent moduli in Eqs.~\eqref{eq:mu_eq}, \eqref{eq:mu_neq} and \eqref{eq:kv}, accounting for experimentally observed property reductions with increasing moisture content and demonstrating the critical importance of considering environmental conditions in composite design and certification.

\begin{figure}[H]
\centering
\begin{subfigure}[b]{0.5\linewidth}
	\centering
	\includegraphics{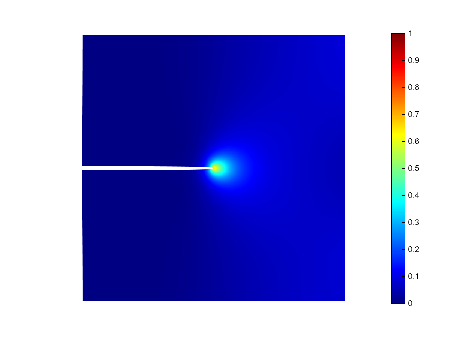}
	\caption{ }
	\label{fig:fracture_randdegree1}
\end{subfigure}
\begin{subfigure}[b]{0.5\linewidth}
	\centering	
	\includegraphics{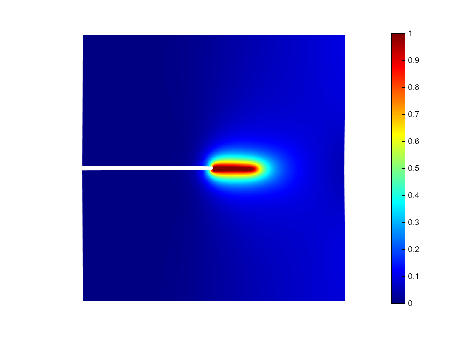}
	\caption{}			
	\label{fig:fracture_randdegree2}
\end{subfigure}	
\begin{subfigure}[b]{0.5\linewidth}
	\centering	
	\includegraphics{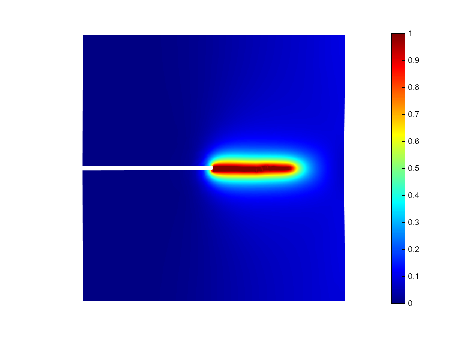}
	\caption{}			
	\label{fig:fracture_randdegree3}
\end{subfigure}%
\begin{subfigure}[b]{0.5\linewidth}
	\centering	
	\includegraphics{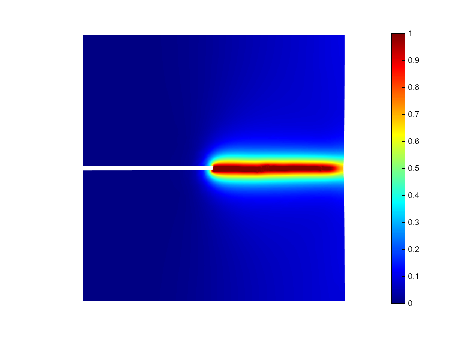}
	\caption{}			
	\label{fig:fracture_randdegree4}
\end{subfigure}
\caption{Crack propagation in randomly distributed GF(50 wt\%)/epoxy at displacements: (a) 0.015~mm, (b) 0.0165~mm, (c) 0.017~mm, and (d) 0.0173~mm. The snapshots are marked at the force-displacement response in Fig.~5.}
\label{fig:damagecontour_randdegree}
\end{figure}

Progressive damage evolution is visualized in Figs.~\ref{fig:fracture_randdegree1}--\ref{fig:fracture_randdegree4}, showing four snapshots of phase-field variable distribution in randomly distributed GF(50 wt\%)/epoxy at displacements 0.015, 0.0165, 0.017, and 0.0173~mm, spanning from pre-peak loading through post-peak softening as shown by four markers in Fig.~\ref{fig:forcedisp_rand-degree_moisture}. At the earliest stage corresponding to approximately the peak load, a small localized damage zone initiates at the notch tip with phase-field variable reaching only moderate values ($\phi \approx 0.2$-$0.3$), indicating stress concentration without significant crack advance. Slightly past peak load, a fully developed crack propagates horizontally across the specimen width, maintaining a straight path perpendicular to loading consistent with mode-I fracture, though minor irregularities reflect local interactions with random fiber distribution. Well into post-peak regime, the crack has propagated nearly to the opposite edge, leaving only a small ligament. This sequence demonstrates that crack propagation in randomly oriented short fiber-reinforced composites is similar to idealized straight cracks in isotropic materials.

\begin{figure}[H]
\centering
\includegraphics{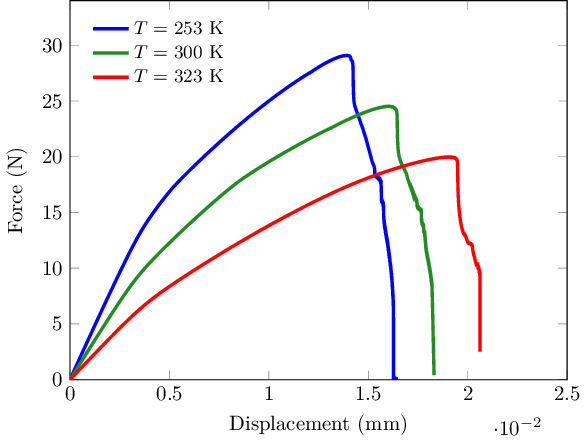}
\caption{Effect of temperature on the force--displacement response of randomly distributed GF (50 wt\%)/epoxy at moisture-saturated condition.}
\label{fig:forcedisp_temp}
\end{figure}

Temperature dependence of viscoelastic-viscoplastic behavior is examined in Fig.~\ref{fig:forcedisp_temp} for randomly distributed GF(50 wt\%)/epoxy under saturated conditions at 253, 300, and 323~K. Results reveal strong temperature sensitivity consistent with thermal activation processes embedded in the constitutive model. At the lowest temperature, the composite exhibits highest stiffness and peak force (approximately 29~N at 0.015~mm), reflecting reduced molecular mobility and increased resistance to viscous flow, with relatively brittle behavior and sharp post-peak softening indicating limited energy dissipation through viscous mechanisms. At room temperature, intermediate stiffness and peak force (approximately 25~N at 0.016~mm) are observed with more gradual post-peak softening, representing a transition regime where both elastic and viscous mechanisms contribute significantly. At elevated temperature, noticeably reduced stiffness and lowest peak force (approximately 20~N at 0.017~mm) are observed, but displacement at failure increases substantially demonstrating enhanced ductility, with much more gradual post-peak behavior and extended tail indicating significant energy dissipation through thermally activated viscous processes. The temperature dependence arises from the modified Kitagawa model (Eqs.~\eqref{eq:mu_eq} and \eqref{eq:mu_neq}) for shear moduli and Argon viscous flow rate (Eq.~\eqref{eq:eps_dot_v}), ensuring that as temperature increases, the matrix becomes softer and viscous relaxation accelerates, leading to observed trends. The approximately 30\% reduction in peak load from 253 to 323~K demonstrates the critical importance of temperature compensation in composite structures operating over wide environmental ranges, and suggests the model could be extended to predict creep and stress relaxation at elevated temperatures—phenomena of practical importance for long-term structural integrity assessments.

\begin{figure}[H]
\centering
\includegraphics{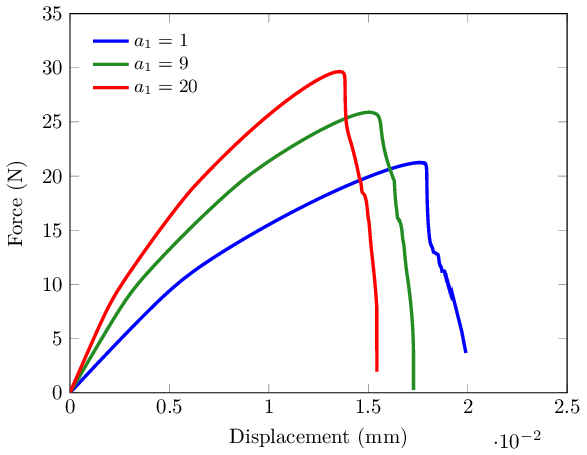}
\caption{Effect of fiber performance on the force--displacement response of randomly distributed GF (50 wt\%)/epoxy at 300~K and dry condition.}
\label{fig:forcedisp_fiberperform}
\end{figure}

The influence of fiber mechanical properties is investigated through parameter $a_1$ in the fiber strain energy function (Eq.~\eqref{eq:f_function}), which controls the initial stiffness contribution of fiber families. Figure~\ref{fig:forcedisp_fiberperform} compares three values: $a_1 = 1$ (low fiber stiffness), $a_1 = 9$ (baseline), and $a_1 = 20$ (high fiber stiffness) for randomly distributed GF(50 wt\%)/epoxy at 300~K under dry conditions, with fiber volume fraction and orientation distribution held constant to isolate intrinsic fiber property effects. Increasing $a_1$ systematically enhances both initial stiffness and peak load: from approximately 21~N at 0.018~mm for $a_1 = 1$, to 25~N at 0.016~mm for $a_1 = 9$, and 30~N at 0.015~mm for $a_1 = 20$. The force-displacement curves diverge immediately from loading onset, confirming that $a_1$ directly affects elastic response rather than just influencing damage evolution or post-peak behavior. Interestingly, post-peak softening also depends on $a_1$: higher fiber stiffness leads to more abrupt failure after peak load, suggesting more brittle fracture behavior attributable to increased stress concentration at fiber ends when stiffer fibers carry higher loads, promoting localized matrix cracking and fiber-matrix debonding. The practical implication is that while using higher-modulus fibers increases strength and stiffness, it may reduce damage tolerance and fracture toughness if fiber-matrix interface properties are not correspondingly improved. These simulations demonstrate the model's capability to assess trade-offs between different reinforcement choices and guide material selection for applications requiring either high strength (favor high $a_1$) or high toughness (favor moderate $a_1$ with good interfacial adhesion).

The crack propagation through layered composite architectures with varying fiber orientations--a configuration widely employed in structural laminates--is investigated in the following figures. Figs.~\ref{fig:forcedisp_layered1} and~\ref{fig:damagecontour_layered1} examine bi-layer GF(50 wt\%)/epoxy at 300~K under dry conditions where top and bottom halves have distinct fiber alignment, investigating how the anisotropy parameter $\hat{\alpha}$ influences crack path deflection at layer interfaces by comparing $\hat{\alpha} = 2$ and $\hat{\alpha} = 10$. The force-displacement response shows that increasing anisotropy significantly enhances both peak load (from approximately 25~N to 28~N) and displacement at peak load (from 0.015~mm to 0.018~mm). More importantly, post-peak behavior differs dramatically: for $\hat{\alpha} = 2$, catastrophic failure occurs shortly after peak with steep load drop, whereas $\hat{\alpha} = 10$ exhibits more gradual softening with a pronounced plateau region around 27~N extending over approximately 0.002~mm displacement range. This plateau indicates stable crack propagation through the layered structure, with the crack repeatedly deflecting and reorienting as it encounters layer interfaces with different fiber orientations—a phenomenon unique to layered composites with strong anisotropic toughness.

\begin{figure}[H]
\centering
\includegraphics{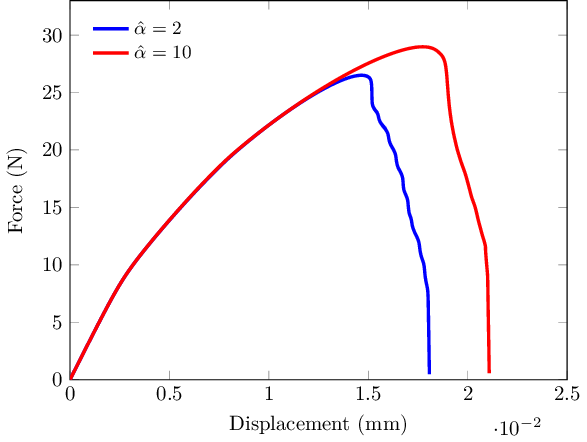}
\caption{Force--displacement response of a layered GF (50 wt\%)/epoxy composite at 300~K and dry condition.}
\label{fig:forcedisp_layered1}
\end{figure}

\begin{figure}[H]
\centering
\begin{subfigure}[b]{0.5\linewidth}
	\centering
	\includegraphics{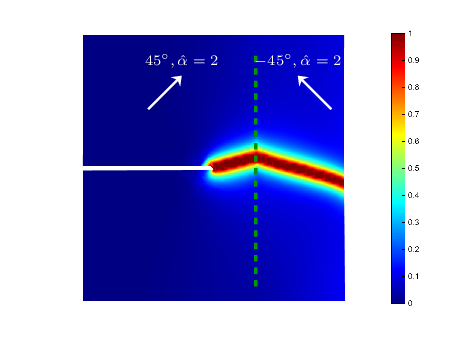}
	\caption{ }
	\label{fig:fracture_layered1_1}
\end{subfigure}%
\begin{subfigure}[b]{0.5\linewidth}
	\centering	
	\includegraphics{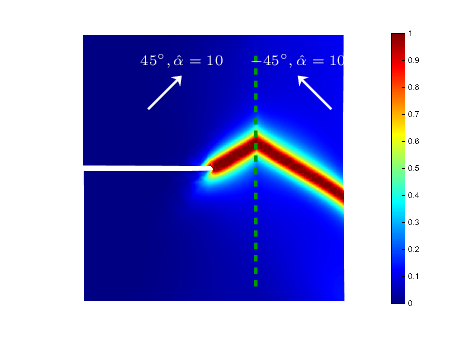}
	\caption{}			
	\label{fig:fracture_layered1_2}
\end{subfigure}	
\caption{Crack propagation in the layered GF(50 wt\%)/epoxy composite: (a) $\hat{\alpha}=2$, and (b) $\hat{\alpha}=10$. The dashed green line indicates the layer interface.}
\label{fig:damagecontour_layered1}
\end{figure}

\begin{figure}[H]
\centering
\includegraphics{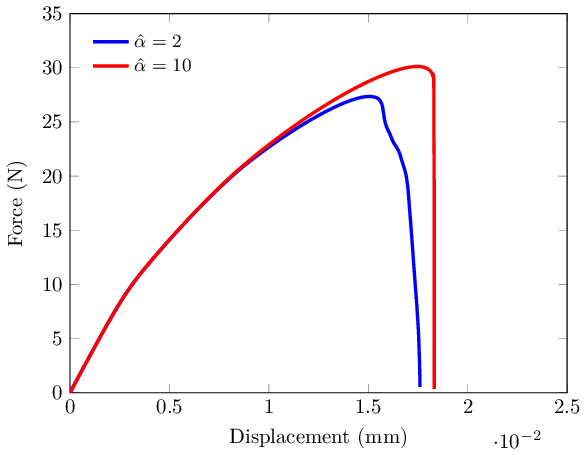}
\caption{Force--displacement response of a layered GF (50 wt\%)/epoxy composite at 300~K and dry condition.}
\label{fig:forcedisp_layered2}
\end{figure}

\begin{figure}[H]
\centering
\begin{subfigure}[b]{0.5\linewidth}
	\centering
	\includegraphics{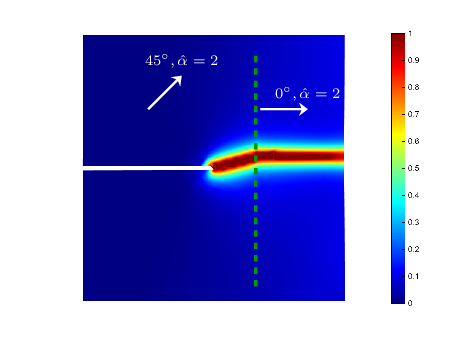}
	\caption{ }
	\label{fig:fracture_layered2_1}
\end{subfigure}
\begin{subfigure}[b]{0.5\linewidth}
	\centering	
	\includegraphics{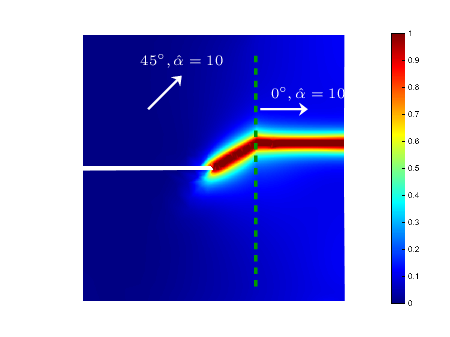}
	\caption{}			
	\label{fig:fracture_layered2_2}
\end{subfigure}	
\caption{Crack propagation in the layered GF(50 wt\%)/epoxy composite: (a) $\hat{\alpha}=2$, and (b) $\hat{\alpha}=10$. The dashed green line indicates the layer interface.}
\label{fig:damagecontour_layered2}
\end{figure}

The crack path visualizations reveal underlying mechanisms. For $\hat{\alpha} = 2$ (Fig.~\ref{fig:fracture_layered1_1}), the crack propagates in a relatively smooth path showing only modest deflection when crossing the layer interface, maintaining an approximately horizontal trajectory as weak anisotropy allows penetration through layers without significant resistance from fiber orientation mismatch. In contrast, for $\hat{\alpha} = 10$ (Fig.~\ref{fig:fracture_layered1_2}), the crack path exhibits pronounced deflection, curving upward in the top layer and downward in the bottom layer to align with the preferred propagation direction dictated by fiber orientation in each layer. A second layered configuration examined in Fig.~\ref{fig:forcedisp_layered2} and~\ref{fig:damagecontour_layered2} demonstrates similar trends with moderate deflection for $\hat{\alpha} = 2$ and substantial deviation including possible crack arrest and reinitiation for $\hat{\alpha} = 10$. Notably, in the high anisotropy case, the crack appears to temporarily arrest at the layer interface before reinitiating in the second layer, resulting in discontinuous damage patterns. This crack arrest phenomenon contributes to the force plateau as applied load must increase to overcome additional energy barrier for crack reinitiation across interfaces.

\section{Summary and conclusions}
\label{sec:summary}

This work presented a comprehensive phase-field framework for modeling anisotropic viscoelastic-viscoplastic fracture in short fiber-reinforced polymer composites under hygrothermal environments at finite deformation. The framework unifies several critical aspects traditionally treated separately: anisotropic crack propagation induced by fiber orientation, rate-dependent inelastic deformation through coupled viscoelastic-viscoplastic behavior, and environmental degradation through moisture- and temperature-sensitive material parameters.

In the phase-field fracture formulation, the anisotropic mechanical response from multiple fiber families is captured through second-order orientation tensor decomposition, reducing complex multi-fiber distributions to principal fiber families characterized by eigenvectors and eigenvalues. Hygrothermal effects are incorporated through moisture-induced swelling, thermal expansion, and environment-dependent material parameters. Numerical investigations on glass fiber-reinforced epoxy composites demonstrate the framework's capabilities:


\begin{itemize}
\item The crack driving force decomposition reveals fiber orientation fundamentally governs spatial energy distribution. Temporal evolution shows approximately 25\% reduction in total driving force over relaxation timescales, demonstrating critical time-dependent influences.

\item Balanced fiber distributions (50-50) produce straight crack propagation with higher peak loads and brittle failure, while unbalanced distributions (70-30) cause substantial deflection with reduced strength but enhanced ductility. The anisotropy parameter $\hat{\alpha}$ exhibits systematic relationships with crack deflection, with increases from 1 to 10 progressively enhancing deflection angles, and fracture energy.

\item Moisture absorption causes 19\% strength degradation for 10 wt\% composites and 7\% for 50 wt\% composites, with plasticization producing gradual softening. Temperature variation from 253 K to 323 K produces approximately 30\% peak load reduction, with elevated temperatures increasing ductility through enhanced viscous dissipation.

\item Strong anisotropy ($\hat{\alpha} = 10$) in bi-layer laminates produces crack deflection and layer reinitiation.
\end{itemize}

The framework advances predictive modeling of SFRPs by providing: (1) unified multi-physics coupling within a phase-field fracture model, (2) decomposition of rate-dependent fracture mechanisms, and (3) quantitative microstructure-property relationships through orientation tensor formalism. Future investigations should address explicit fiber-matrix interface modeling through cohesive phase-field formulations, multiscale homogenization incorporating manufacturing-induced fiber distributions, and fatigue damage under cyclic loading. From a methodological standpoint, the unified variational damage model of Ren et al.~\cite{ren2024variational} offers a length scale insensitive phase-field formulation that could improve the objectivity of crack predictions with respect to mesh and regularization parameters in the current viscoelastic–viscoplastic setting. Furthermore, the adaptive deep energy minimization approach~\cite{goswami2020transfer} presents a compelling alternative to the staggered Newton–Raphson solver employed here, potentially reducing computational cost through physics-informed neural network-based adaptivity in the phase-field solution.

\section*{Acknowledgements}
The authors acknowledge the resources and support provided by the Norwegian Research Infrastructure Services (NRIS). The computational work was performed on resources provided through the Sigma2 national e-infrastructure, project NN10041K.

\bibliography{mybibfile}
\appendix
\end{document}